\documentclass{aa}
\usepackage{epsfig}
\usepackage{lscape}
\usepackage{aalongtable}
\usepackage{natbib}
\usepackage{graphics}
\usepackage{txfonts}
\usepackage{color}
\usepackage{times}
\newcommand\ra[4]{~~#1$^{\rm h}$#2$^{\rm m}$#3$^{\rm s}$.#4 }
\newcommand\dec[3] {$#1$$^{\circ}$#2$^{\rm '}$#3$^{\rm ''}$}

\newcommand{\sbb}{mag/$\sq\arcsec$}

\def\h2{H{\small~II}}

\def\upperspace{\rule[0.0ex]{0cm}{2.5ex}}

\newfont{\lx}{cmssdc10 scaled 900}
\def\apj{ApJ}
\def\apjs{ApJS}
\def\aap{A\&A}

\def\aj{AJ}
\def\mnras{MNRAS}

\begin{document}

\title{An investigation of the luminosity-metallicity relation 
for a large sample of low-metallicity emission-line galaxies
\thanks{Based on observations
collected at the European Southern Observatory, Chile, VLT and 3.6m 
telescopes.}, \thanks{Tables 1 - 6 and Figures 1 - 2 are only available in 
electronic form in the online edition.}}

\author{N. G.\ Guseva \inst{1,2}
\and P.\ Papaderos \inst{3,4}
\and H. T.\ Meyer \inst{5,6}
\and Y. I.\ Izotov \inst{1,2}
\and K. J.\ Fricke \inst{1}}
\offprints{N.G. Guseva, guseva@mao.kiev.ua}
\institute{          Max-Planck-Institute for Radioastronomy, 
                     Auf dem H\"ugel 69,
                     53121 Bonn, Germany
\and                 
                     Main Astronomical Observatory,
                     Ukrainian National Academy of Sciences,
                     Zabolotnoho 27, Kyiv 03680,  Ukraine
\and                 
                     Instituto de Astrof\'{\i}sica de Andaluc\'{\i}a (CSIC),
                     Camino Bajo de Hu\'etor 50, Granada E-18008, Spain
\and
                     Department of Astronomy and Space Physics,
                     Uppsala University, Box 515, SE-75120 Uppsala, Sweden
\and
                     Astronomisches Rechen-Institut am Zentrum f\"ur Astronomie (ZAH),
                     M\"onchhofstr. 12-14, 69120 Heidelberg, Germany
\and
                     Institute for Astrophysics, University of 
                     G\"ottingen, Friedrich-Hund-Platz 1, 
                     37077 G\"ottingen, Germany
}

\date{Received \hskip 2cm; Accepted}

\abstract
{We present 8.2m VLT spectroscopic observations of 28 H {{\sc ii}} regions 
in 16 emission-line galaxies and 3.6m ESO telescope spectroscopic observations 
of 38 H {{\sc ii}} regions in 28 emission-line galaxies.
These emission-line galaxies were selected mainly from
the Data Release 6 (DR6) of the Sloan Digital Sky Survey (SDSS) 
as metal-deficient galaxy candidates. 
}
{We collect photometric and high-quality spectroscopic data for a 
large uniform sample of star forming galaxies including new observations. 
Our aim is to study the luminosity-metallicity ($L-Z$) relation  
for nearby galaxies, especially at its low-metallicity end
 and compare it with that for higher-redshift galaxies.
} 
{Physical conditions and element abundances in the new sample are derived 
with the $T_{\rm e}$-method, excluding six H {{\sc ii}} regions from the VLT 
observations and nearly two third of the
H {{\sc ii}} regions from the 3.6m observations. 
Element abundances for the latter galaxies were derived with the 
semiempirical strong-line method. 
}
{
From our new observations we find  that the oxygen abundance in 
61 out of the 66 H {{\sc ii}} regions of our sample
ranges from 12 + log O/H = 7.05 to 8.22. Our 
sample includes 27 new galaxies with 12 + log O/H $<$ 7.6 which 
qualify as extremely metal-poor star-forming galaxies (XBCDs). 
Among them are 10 H {{\sc ii}} regions with 12 + log O/H $<$ 7.3.
The new sample is combined with a further 93 low-metallicity galaxies 
with accurate oxygen abundance determinations from our previous studies, 
yielding in total a high-quality spectroscopic data set
of 154 H {{\sc ii}} regions. 9000 more galaxies with oxygen abundances, based 
mainly on the $T_{\rm e}$-method, are compiled from the SDSS.
Photometric data for all galaxies of our combined sample are 
taken from the SDSS database while distances are from the NED.
Our data set spans a range of 8 mag with respect to its 
absolute magnitude in SDSS $g$ 
(--12 $\ga M_g \ga$ --20) and nearly 2 dex in its 
oxygen abundance (7.0$\la$12 + log O/H$\la$8.8), allowing us to probe 
the $L-Z$ relation in the nearby universe down to the lowest currently
studied metallicity level. 
The $L-Z$ relation established on the basis of the present sample is
consistent with previous ones obtained for
emission-line galaxies.
} 
{}
\titlerunning{The L-Z relation for 
a large sample of low-metallicity galaxies}
\maketitle

\section{Introduction \label{intro}}

It was shown more than 20 years ago that low-luminosity dwarf galaxies
have systematically lower metallicities compared to more luminous galaxies 
\citep{Lequeux1979,Skillman1989,RicherMcC1995}. 
This dependence, initially obtained for irregular galaxies, was 
later confirmed for galaxies of different morphological types 
\citep[e.g.][]{Vila1992,KobylZarit1999,MelbourneSalzer2002,Lee2004,
Pil2004,Lee45mu2006}.

The differences between giant and dwarf galaxies are usually attributed to
different chemical evolution of galaxies with different masses 
\citep[e.g. ][]{Lequeux1979,
Tremonti2004,Lee45mu2006,Ellison2008,Gavilan2009}. 
Thus, more efficient mechanisms seem to be at work in massive galaxies 
converting gas into stars and/or less efficient ones 
ejecting enriched matter into the galactic halo or even 
into the intergalactic medium.
While the mass of a galaxy is one of the key physical 
parameters governing galaxy evolution, 
its determination is not easy and somewhat uncertain. 
Therefore, very often the luminosity, which is directly derived from 
observations, is used instead of the mass. 
In addition, some authors also use
other global characteristics of a galaxy 
such as Hubble morphological type, rotation velocity, the gas mass fraction, 
surface brightness of the galaxy, to study correlations between metallicity 
and macroscopic properties of a galaxy \citep[e.g. ][]{Tremonti2004,Pil2004}.

Metallicity reflects the level of the gas astration in the galaxy.
Hence, the metallicity of a galaxy depends strongly on its evolutionary 
state, specifically, on the fraction of the gas converted into stars.
The metallicity in emission-line galaxies is defined in terms of the 
relative abundance of oxygen to hydrogen (usually 12 + log O/H) in the 
interstellar medium (ISM).
Different mechanisms were considered in chemical evolution models to account 
for the low metallicity of dwarf galaxies, mainly 1) enriched galactic wind 
outflow which expells the newly synthesized heavy elements from the galaxy,
resulting in slowing enrichment of the galaxy ISM;
2) inflow of metal-poor intergalactic gas and its mixing with the galaxy ISM
which results in decreasing ISM metallicity, and 3) the burst character
of star formation with a very low level of astration between the bursts.
In principle, chemical evolution models could predict the slope and scatter 
of the mass-metallicity $M-Z$ (and luminosity-metallicity $L-Z$) 
relations over a large range in mass (luminosity) and metallicity
invoking the mechanisms mentioned above.  

Usually, $L-Z$ relations are based on optical observations of nearby
galaxies. However, it was shown in recent studies that the near infrared 
(NIR) range could be more promising for such studies.
\citet{Saviane2008} collected abundances obtained by means of the 
temperature-sensitive method  
and NIR luminosities for a sample of dwarf irregular galaxies 
with --20 $<$ $M_H$ $<$ --13, located in nearby groups of galaxies.
They obtained a tight $M-Z$ relation with 
a low scatter of 0.11 dex around its linear fit.
\citet{Salzer05} \citep[see also ][]{Vaduvescu07} 
noted that the NIR luminosities are more fundamental than the 
$B$-band ones, since they are largely free of absorption effects   
and are more directly related to the stellar mass of the galaxy than
optical luminosities.
Nevertheless, this statement is correct only for galaxies with low 
and moderate SF activity. 
In emission-line galaxies with high star formation rate (SFR), such as 
blue compact dwarf (BCD) galaxies, the young, 
low mass-to-light ($M/L$) ratio stellar component may provide up
to $\sim$50\% of the total $K$ band emission \citep{Noeske03}.
Additionally, in such systems the contribution of ionized gas to the total 
luminosity could be high \citep[see e.g.][]{I97b,P98,P02}, especially in the 
NIR range \citep[see e.g. ][]{Vanzi00,Smith2009}, and should be taken
into account.

Recently, studies of the $L-Z$ relation were extended 
to larger volumes by including moderate- and high-redshift galaxies 
\citep{KobylZarit1999,Contini2002,Maier2004}. 
Variations of the $L-Z$  relation with redshift can provide 
a means to study the galaxy evolution with look-back time 
\citep[see, e.g., ][]{Kobulniky2003}. 
It was established in this study that the slopes and zero points of the $L-Z$ 
relation evolve smoothly with redshift. 
Its large dispersion has been attributed to galaxy evolution effects.
However, these results and their comparison with those for nearby
galaxies should be considered with caution. The high-redshift samples 
are biased by different selection criteria and metallicity
calibrations as compared to the local galaxies. 
They consist on average of more luminous 
and higher metallicity galaxies. 
Star-forming dwarf galaxies in the relatively 
high-redshift (up to $z$ $\sim$ 1) samples are rare 
because of their intrinsic faintness. 
Moreover, due to the weakness of the [O {\sc iii}]$\lambda$4363 emission 
line in the spectra of these galaxies,
their abundance determinations are more uncertain and 
could lead to a large scatter in the $L-Z$ diagrams. 
This fact could be the reason for a larger scatter
of high-redshift dwarf galaxies if the direct $T_{\rm e}$-method 
is used instead of the empirical R$_{23}$ one \citep[e.g., ][]{Kakazu07}. 

In summary, it is difficult to obtain reliable metallicities over a large
luminosity range in a homogeneous manner, i.e. employing a unique technique
(e.g. the direct $T_{\rm  e}$-method), even for nearby galaxies.
Therefore, different methods for abundance determination are applied for
galaxies of different types. 
The direct method is mainly used for nearby low-metallicity galaxies, 
while various empirical methods are used for nearby high-metallicity 
galaxies and for almost all high-redshift galaxies.
The variety of methods results in significant differences 
in the $L-Z$ relations obtained with the direct $T_{\rm e}$-method and 
those based on strong emission line ratio calibrations, 
such as $R_{23}$, $P$-method, N2 and O3N2 methods.
These differences were reported by many authors 
\citep[e.g., ][]{Pil2004,Shi2005,Hoyos2005,Kakazu07}.

Large surveys, such as the Two-Degree Field Galaxy Redshift Survey (2dFGRS)
and Sloan Digital Sky Survey (SDSS), provide rich data sets for 
statistically improved studies of the $L-Z$ relation.
For example, \citet{Lamareille2004} using more than 6000 spectra of
SF galaxies at $z$ $<$ 0.15 from the 2dFGRS have obtained an $L-Z$ 
relation that is much steeper than that for nearby irregulars and spiral 
galaxies. 
\citet{Tremonti2004} studied the mass-metallicity ($M-Z$) relation for
53000 SF galaxies within $z$ $\sim$ 0.2 extracted from 
SDSS, using their stellar continuum and line fitting method.
This method is applicable because the bulk of their 
emission-line galaxies show weak emission lines and strong 
stellar absorption features, and therefore the contribution of gaseous 
emission to the galaxy luminosity is low.
The \citet{Tremonti2004} $M-Z$ relation is relatively steep but it flattens 
for massive galaxies at masses above 10$^{10}$ M$_{\odot}$.
On the contrary, \citet{MelbourneSalzer2002} using 519 emission-line galaxies 
from the KPNO International Spectroscopic Survey (KISS) found that 
the slope of the $L-Z$ relation for luminous galaxies
is steeper than that for dwarf galaxies.
Nevertheless, \citet{Pil2004} have compared the $L-Z$ relation based 
on more than 1000 published spectra of H {\sc ii} regions in spiral
galaxies to that for irregular galaxies. 
They found that the slope of the relation for spirals is slightly 
shallower than the one for irregular galaxies.
Furthermore, using 72 star-forming galaxies, \citet{Shi2005}
have also shown that the slope of the $L-Z$ relation for luminous 
galaxies is slightly shallower than that for dwarf galaxies.

Is the slope of the $L-Z$ relation invariant for galaxies of 
different type, such as local dwarf and spiral galaxies and 
high-redshift galaxies? 
If differences in the $L-Z$ relations for intermediate- and 
high-redshift and local ones are present, they may yield
important constraints on the Star Formation History (SFH) 
of galaxies.
For this, an as accurate as possible $L-Z$ relation, 
based on homogeneous 
high-quality photometric and spectroscopic data is required for galaxies 
that covers a large range in metallicity and luminosity.
In particular, probing the slope of the $L-Z$ relation in its  
low-metallicity end, i.e. in the range expected for unevolved low-mass
galaxies in the faraway universe, is much needed.   
For this purpose, in this paper we focus our study 
on the lowest-metallicity galaxy candidates selected from large 
spectroscopic surveys, using deep follow-up spectroscopic observations.
   
Specifically, the objective of the work is to study 
the $L-Z$ relation for a large uniform sample of emission-line galaxies in 
the local Universe for which the element abundances are obtained with 
high precision.
The main feature of our sample is that it is one of the richest currently
available at the low-metallicity end.
For the galaxy selection we used different surveys such as 2dFGRS, SDSS and 
others. Most of our sample galaxies currently undergo strong episodes of 
star formation (SF).

\setcounter{figure}{2}
\begin{figure*}[t]
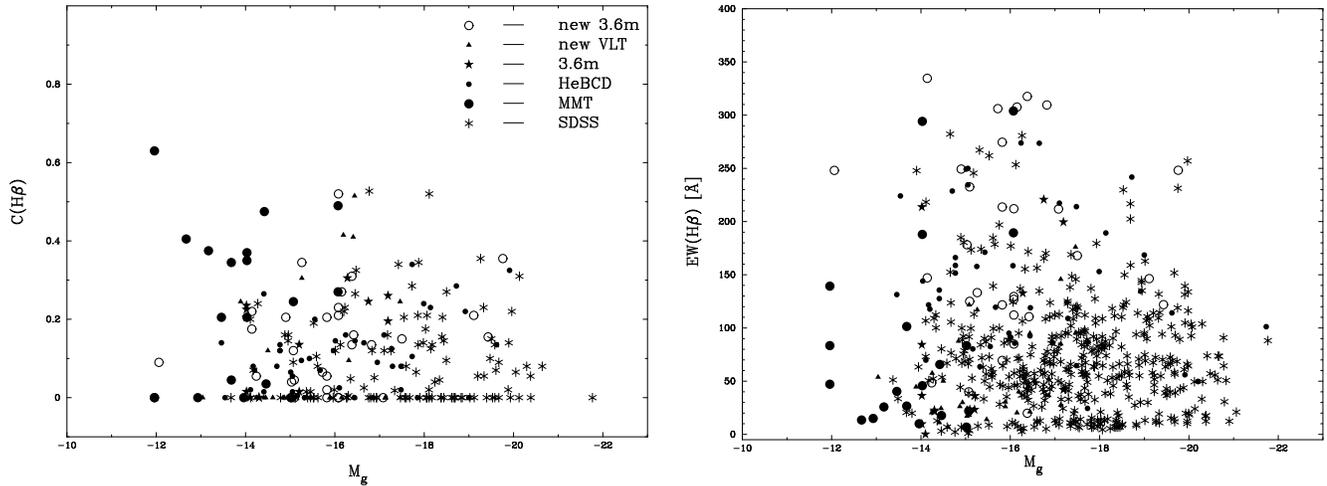

\hspace*{0.0cm}\psfig{figure=12414f3a.ps,angle=-90,width=8.5cm,clip=}
\hspace*{0.4cm}\psfig{figure=12414f3b.ps,angle=-90,width=8.5cm,clip=}
\caption{The reddening parameter $C$(H$\beta$) obtained from the observed
Balmer decrement (left) and the logarithm of the H$\beta$ 
equivalent width (in $\AA$) (right) as a function of the 
absolute magnitude $M_g$ in the SDSS $g$-band.
Oxygen abundances are derived mainly with the $T_{\rm e}$-method. 
Additionally, only high-quality SDSS spectra 
with the non-detected [O {\sc iii}]$\lambda$4363 emission line are included
for which oxygen abundances are derived with the semiempirical method
\citep{IT2007}.
New 3.6m ESO telescope and VLT data are shown by open circles and by stars, 
respectively.
The additional sample of the metal-poor galaxies collected from previous  
3.6m ESO observations \citep{BJlarge2007} are shown by filled triangles.
Filled circles denote the data from the HeBCD sample collected 
by \citet{ISGT2004} and \citet{IT04a}. 
The MMT data from \citet{IT2007} are shown by large filled circles.
The comparison SDSS sample is represented by asterisks in which are excluded
the H {\sc ii} regions in nearby spiral galaxies, faint galaxies 
with $m_g$ $>$ 18, 
the nearest galaxies with redshift $z$ $<$ 0.004 and all galaxies with 
$\sigma$$I$(4363)/$I$(4363) $>$ 0.25, resulting in the SDSS sample of 443 objects.
}
\label{fig3}
\end{figure*}

We performed 3.6m ESO spectroscopic observations 
of a sample of 38 H {\sc ii} regions in 28 emission-line galaxies and 
8.2m VLT spectroscopic observations of a sample of 28 H {\sc ii} regions 
in 16 emission-line galaxies.
We supplement our new data with our previous data collected from the MMT 
observations \citep{IT2007}, 
and from the 3.6m ESO observations \citep{BJlarge2007,Pap2008}
of the low-metallicity emission-line galaxies selected from the SDSS,
with the sample used by \citet{IT04a} to study the helium abundance 
in low-metallicity BCDs (henceforth referred to as the HeBCD sample) 
and with the MMT sample used by \citet{TI2005} to 
study high-ionization emission lines in low-metallicity BCDs.
Our MMT, 3.6m ESO and HeBCD low-metallicity galaxies were
selected from different surveys 
\citep[a more complete description of the MMT, 3.6m ESO
and HeBCD subsamples can be found in ][]{IT2007,BJlarge2007,Pap2008,IT04a}. 
During past years we selected from the SDSS and performed follow-up 
spectroscopic observations with large telescopes of (i) BCDs with strong 
ongoing SF, i.e. galaxies with high EW(H$\beta$), blue colours, high ionisation
parameter, and (ii) low-metallicity galaxies in a relatively 
quiescent phase of SF, i.e. galaxies with low EW(H$\beta$), 
low ionisation parameter or older starburst age.
For this, we selected galaxies with weak or not detected 
[O {\sc iii}]$\lambda$4363 emission line and with 
[O {\sc iii}]$\lambda$4959/H$\beta$ $\la$ 1 and 
[N {\sc ii}]$\lambda$6583/H$\beta$ $\la$ 0.05 \citep{IPGFT2006,IT2007}.

SDSS is an excellent source of both photometric and
spectroscopic data for more than one million galaxies in its
Data Release 7 (DR7) \citep{A09}. 
Despite that, our stringent selection criteria 
resulted in a very small sample of extremely metal-deficient emission-line
galaxies with reliable abundance determinations.
This sample is supplemented for the purpose of comparison 
by a sample of $\sim$ 9000 SDSS emission-line galaxies (SDSS sample) 
over a larger range of metallicities. 
The oxygen abundances for the galaxies from the SDSS sample 
are obtained using the direct $T_{\rm e}$-method.
In addition, only high-quality spectra of SDSS galaxies with the 
non-detected 
[O {\sc iii}]$\lambda$4363 emission line are included, for which oxygen 
abundances are derived by a semiempirical method \citep{IT2007}.

Thus, we construct a large homogeneous sample with uniform selection 
criteria, uniform data reduction methods, and uniform techniques 
for the element abundance determinations.
The apparent $g$ magnitudes for our entire data set are taken from the SDSS.
They were used to derive absolute $g$ magnitudes which were 
corrected for the Galactic extinction and Virgo cluster infall, except for
the comparison SDSS sample galaxies. For the latter galaxies the absolute
magnitude was derived from the observed redshift, adopting a
Hubble constant of $H_0$ = 75 km s$^{-1}$ Mpc$^{-1}$. 

 The paper is organized as follows. Observations and data reduction are 
described in Sect. 2. Physical conditions and element abundances in the 
galaxies from the new observations are presented in Sect. 3. We discuss the 
properties of the $L-Z$ relation in Sect. 4 and summarise our conclusions in
Sect. 5.

\section{Observations and data reduction}

The new spectra of the 3.6m ESO sample were 
obtained on 14 - 16 September, 2007 with the spectrograph EFOSC2.
The grism Gr\#7 and a long slit with the width of 1\farcs2
were used yielding a wavelength coverage
of $\lambda$$\lambda$3400--5200\AA.
The long slit was centered on the brightest part of each galaxy and 
simultaneously on another H {\sc ii} regions, whenever present.
The name of each galaxy with its different H {\sc ii} regions, the 
coordinates R.A., Dec. (J2000.0), date of observation, 
exposure time, number of exposures for each observation,
average airmass and seeing are given in Table~\ref{obs36}.  
All spectra were obtained at low airmass or with the slit oriented along the 
parallactic angle,
so no corrections for atmospheric refraction have been applied.

The new VLT spectra were obtained during several runs in October - December, 
2006 and January,  2007 with the spectrograph FORS2 mounted at the ESO VLT
UT2.
The observing conditions were photometric during the nights with seeing
$<$ 1\farcs5. Several observations were performed under excellent 
seeing conditions ($<$ 0\farcs8).
The grisms 600B ($\lambda$$\lambda$$\sim$3400--6200) 
and 600RI and filter GG435 ($\lambda$$\lambda$$\sim$5400--8620) for the blue 
and red parts of the spectrum, respectively, were used.
A 1\arcsec$\times$360\arcsec\ long slit was centered on the brightest 
H {\sc ii} regions of each galaxy.
In Table~\ref{obsVLT}, the same parameters as in Table~\ref{obs36} are given 
for the VLT observations. Note that for each galaxy the first and the second
lines are related to the observations in the blue and red ranges, respectively.
Again, as for the EFOSC2 spectra, the observations were obtained 
at low airmass, and no corrections for atmospheric refraction were applied.

The data were reduced with the IRAF\footnote{IRAF is 
the Image Reduction and Analysis Facility distributed by the 
National Optical Astronomy Observatory, which is operated by the 
Association of Universities for Research in Astronomy (AURA) under 
cooperative agreement with the National Science Foundation (NSF).}
software package. 
This included  bias--subtraction, 
flat--field correction, cosmic-ray removal, wavelength calibration, 
night sky background subtraction, correction for atmospheric extinction and 
absolute flux calibration of the two--dimensional spectrum.
The spectra were also corrected for interstellar extinction using the 
reddening curve of \citet{W58}.
   One-dimensional spectra of one or several H {\sc ii} regions 
in each galaxy were extracted from two-dimensional observed spectra. 
   The flux- and redshift-calibrated one--dimensional EFOSC2 3.6m spectra 
of the H {\sc ii} regions are shown in Fig.~\ref{fig1} for all galaxies 
given in Table~\ref{obs36}.  One-dimensional VLT spectra 
are shown in Fig.~\ref{fig2} for 28 objects 
listed in Table~\ref{obsVLT}. 
 For VLT spectra of the four background galaxies and H {\sc ii} region No.2 in 
the galaxy J2354-0004 without a detectable [O {\sc iii}]$\lambda$4363$\AA$ 
emission line, no abundance determination has been done.

Emission-line fluxes were measured using Gaussian profile fitting. 
The errors of the line fluxes were calculated from the photon statistics
in the non-flux-calibrated spectra. They have been propagated in the 
calculations of the elemental abundance errors. The quality of the
VLT data reduction could be verified by a comparison
of He {\sc i} $\lambda$5876 emission line fluxes measured in the blue and
red spectra of the same object. We found that the fluxes of the 
He {\sc i} $\lambda$5876 emission line in spectra of bright objects
differ by no more than 1-2\% indicating an accuracy in the 
flux calibration at the same level. 
For faint objects the difference between the flux
of the He {\sc i} $\lambda$5876 emission line in the blue and red spectra
is higher, $\sim$5 -- 10\%, and is comparable to the statistical errors 
listed in Table \ref{t4_1_VLT}.

The extinction coefficient 
$C$(H$\beta$) and equivalent widths of the hydrogen absorption lines
EW(abs) are calculated simultaneously, minimizing the deviations 
of corrected fluxes $I(\lambda)$/$I$(H$\beta$) of all hydrogen Balmer lines
from their theoretical recombination values as
\begin{eqnarray*}
\frac{I(\lambda)}{I({\rm H}\beta)} & = &\frac{F(\lambda)}{F({\rm H}\beta)}
\frac{EW(\lambda)+|EW(abs)|}{EW(\lambda)}\frac{EW({\rm H}\beta)}{EW({\rm H}\beta)+|EW(abs)|} \\
                       & \times & 10^{C({\rm H}\beta)f(\lambda)}.
\end{eqnarray*} 
Here $f$($\lambda$) is the reddening function
normalized to the value at the wavelength of the H$\beta$ line,
$F$($\lambda$)/$F$(H$\beta$) are the 
observed  hydrogen Balmer emission line fluxes relative to that
of H$\beta$, EW($\lambda$) and EW(H$\beta$) 
the equivalent widths of emission
lines, and EW(abs) 
the equivalent widths of hydrogen absorption lines which we
assumed to be the same for all hydrogen lines. For $f$($\lambda$) we adopted 
the
reddening law by \citet{W58}. 
The extinction-corrected fluxes of emission lines other than hydrogen ones
are derived from equation 
\begin{eqnarray*}
\frac{I(\lambda)}{I({\rm H}\beta)} & = &\frac{F(\lambda)}{F({\rm H}\beta)}
                       \times 10^{C({\rm H}\beta)f(\lambda)} .
\end{eqnarray*} 

The extinction-corrected emission line fluxes $I$($\lambda$) relative to the 
H$\beta$ fluxes multiplied by 100, the extinction coefficients 
$C$(H$\beta$), the equivalent widths EW(H$\beta$),
the observed H$\beta$ fluxes $F$(H$\beta$) 
(in units 10$^{-16}$ erg s$^{-1}$ cm$^{-2}$), and the 
equivalent widths of the hydrogen absorption lines
are listed in Table \ref{t3_1_36} (3.6m ESO observations) and in
Table \ref{t4_1_VLT} (VLT observations). 
$C$(H$\beta$) and EW(abs) are set to zero in
Tables \ref{t3_1_36} and \ref{t4_1_VLT} 
if we do not have enough observational data or their values are negative.

\begin{figure}
\hspace*{0.0cm}\psfig{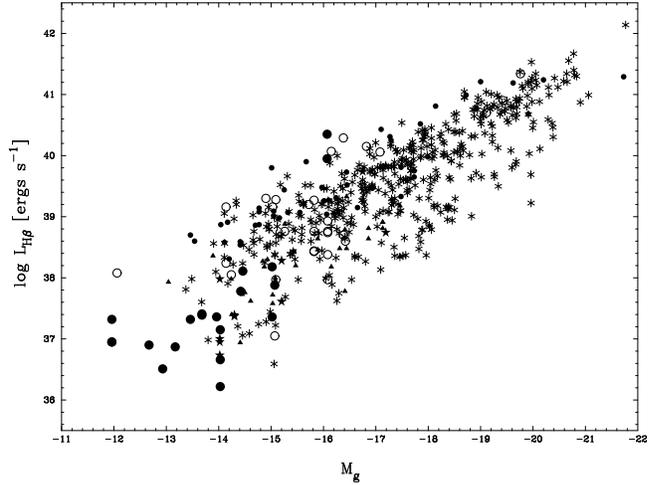}
\caption{The logarithm of the H$\beta$ line luminosity
(in ergs s$^{-1}$) vs absolute magnitude. 
The same samples and symbols as in Fig~\ref{fig3} are used.
}
\label{fig4}
\end{figure}

\begin{figure*}[t]
\hspace*{1.0cm}\psfig{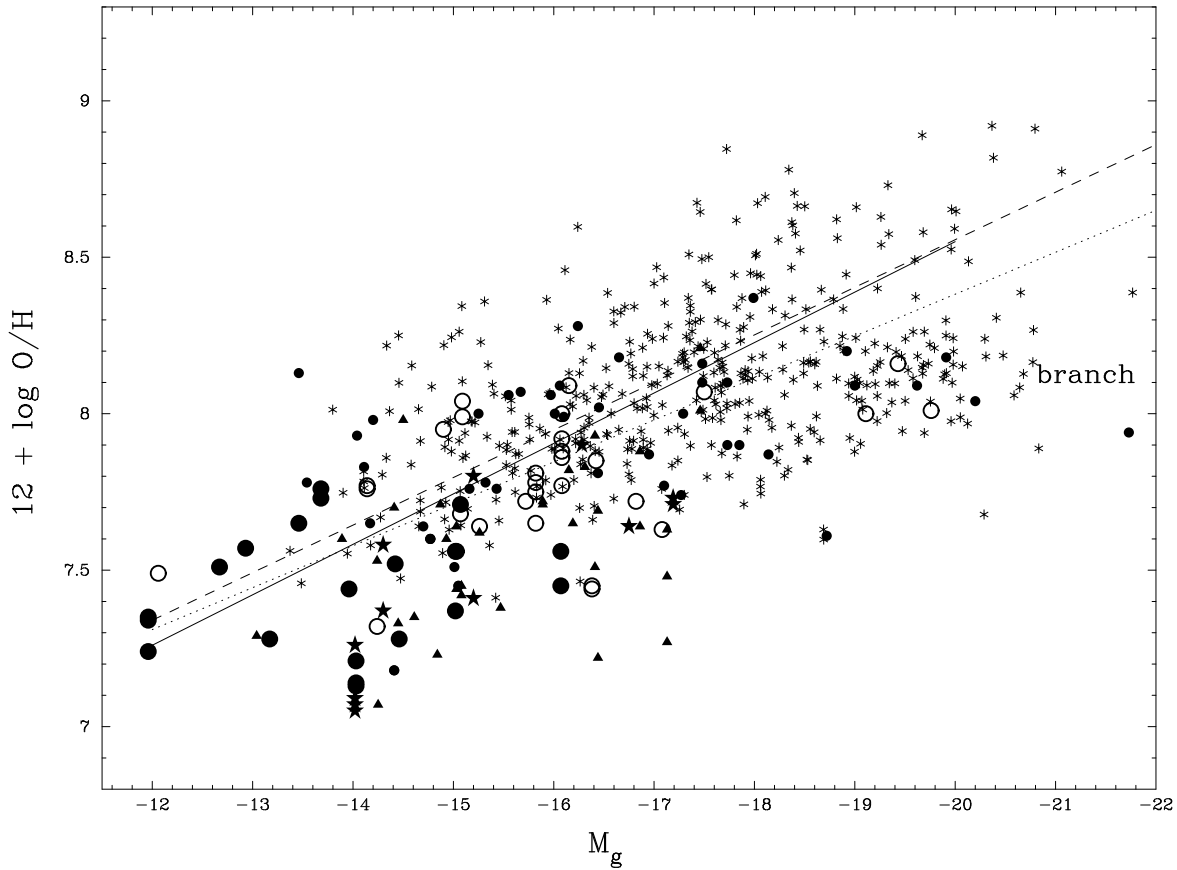}
\caption{Oxygen abundance vs absolute magnitude for a large galaxy
    sample. The same samples and symbols as in Fig~\ref{fig3} are used.
The region denoted as a ``branch'' is populated mainly by the galaxies
with relatively high redshifts ($z$ $>$ 0.02) and oxygen abundances derived by 
the $T_{\rm e}$-method.
The dotted line is a mean least-squares fit to all our data, 
while the solid line is a mean least-squares fit to our data 
excluding ``branch'' galaxies with $M_g$ $<$ --18.4 and oxygen 
abundances in the range 8.0--8.3. 
The dashed line is a mean least-squares fit to the 
local dwarf irregular galaxies by \citet{Skillman1989}.
}
\label{fig5}
\end{figure*}

\begin{figure*}[t]
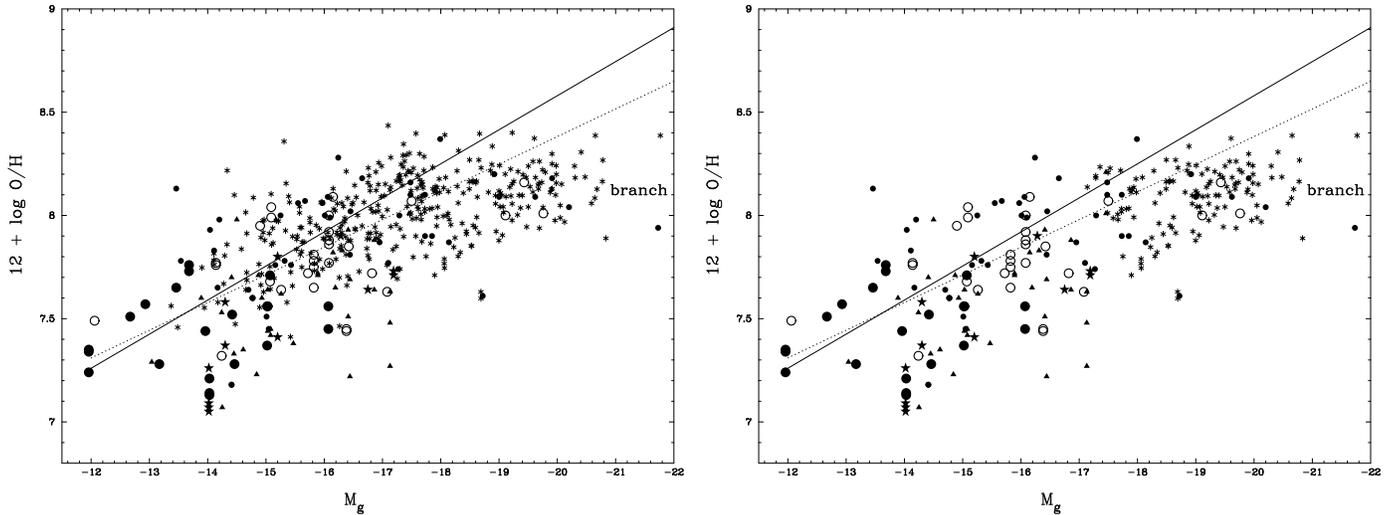

\hspace*{0.2cm}\psfig{figure=12414f6a.ps,angle=-90,width=9.0cm,clip=}
\hspace*{0.2cm}\psfig{figure=12414f6b.ps,angle=-90,width=9.0cm,clip=}
\caption{{The same as in Fig~\ref{fig5} but (left) only SDSS galaxies with
oxygen abundances derived with the $T_{\rm e}$-method and (right) 
only relatively high-redshift galaxies with z$>$ 0.02. 
The dotted line is a mean least-squares fit to all our data, 
while the solid line is a mean least-squares fit to our data 
excluding ``branch'' galaxies.}
}
\label{fig6}
\end{figure*}

\begin{figure*}[t]
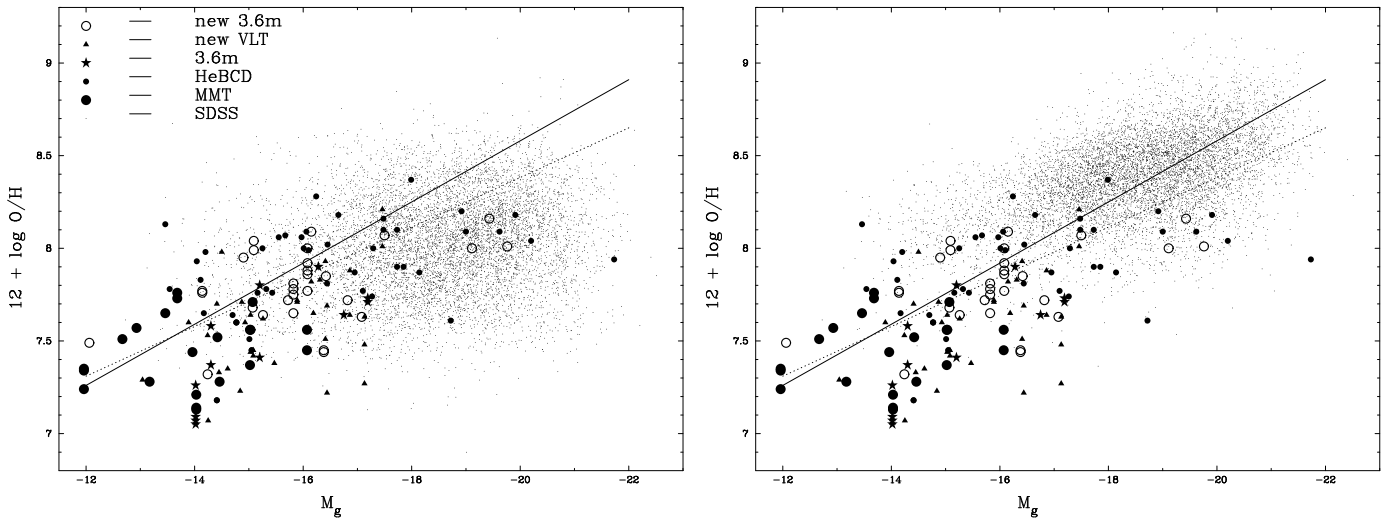

\hspace*{0.2cm}\psfig{figure=12414f7a.ps,angle=-90,width=9.0cm,clip=}
\hspace*{0.2cm}\psfig{figure=12414f7b.ps,angle=-90,width=9.0cm,clip=}
\caption{Oxygen abundance vs absolute magnitude. H {\sc ii} regions in spiral
galaxies and the nearest SDSS galaxies with 
redshift $z$ $<$ 0.004 are excluded from the SDSS sample while faint 
galaxies with 
$m_g$ $>$ 18 are included, resulting in an SDSS sample of 7964 objects. 
Symbols are the same as in Fig.~\ref{fig3} except for SDSS galaxies which
are shown by dots. In the left panel the oxygen abundances 
12 +logO/H for SDSS galaxies are obtained with the $T_{\rm e}$-method.  
In the right panel 12 +logO/H for SDSS galaxies is derived 
with the semiempirical method. 
The dotted line is a mean least-squares fit to all our data 
from Fig.~\ref{fig5}, while the solid line is a mean 
least-squares fit to the same data excluding ``branch'' galaxies.
}
\label{fig7}
\end{figure*}

\begin{figure*}[t]
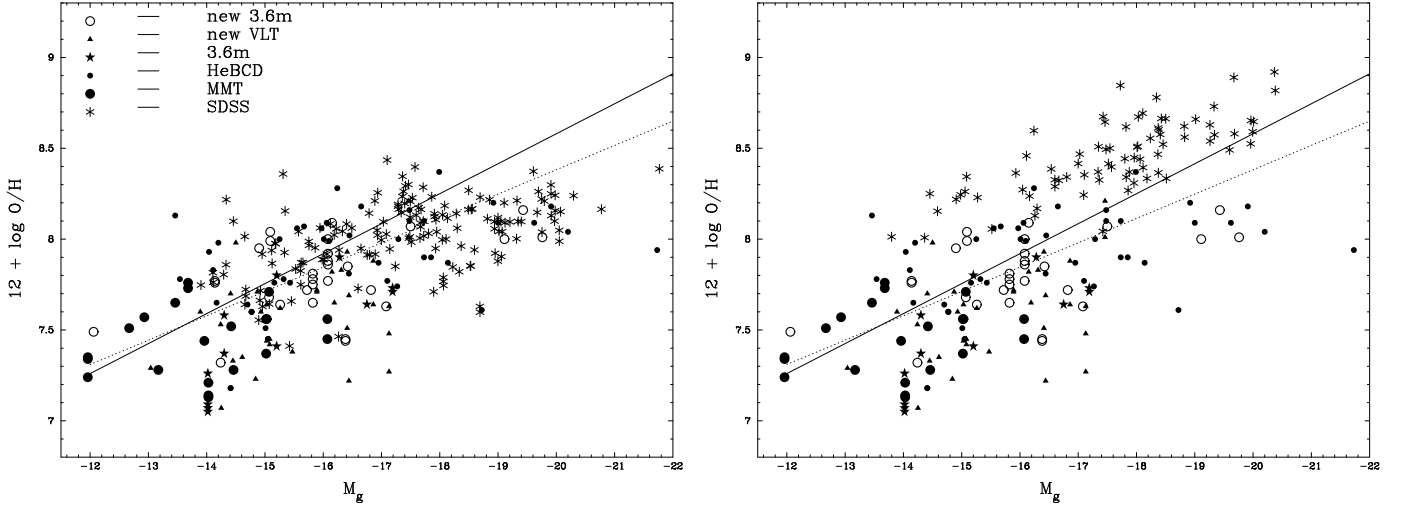

\hspace*{0.2cm}\psfig{figure=12414f8a.ps,angle=-90,width=9.0cm,clip=}
\hspace*{0.2cm}\psfig{figure=12414f8b.ps,angle=-90,width=9.0cm,clip=}
\caption{
The same as in Fig.~\ref{fig5} but 
only SDSS galaxies with EW(H$\beta$) $>$ 80$\AA$ (left panel) and 
with EW(H$\beta$) $<$ 20$\AA$ (right panel).
The dotted line is a 
mean least-squares fit to all our data from Fig.~\ref{fig5}, 
while the solid line is a mean 
least-squares fit to the same data excluding ``branch'' galaxies.
 For the SDSS galaxies with EW(H$\beta$) $>$ 80$\AA$ the oxygen abundances are 
$\sim$ 0.4 dex lower than the ones for the galaxies with 
EW(H$\beta$) $<$ 20$\AA$.
The galaxies with the lowest metallicities could be found more easily 
among the H {\sc ii} regions with high EW(H$\beta$).
}
\label{fig8}
\end{figure*}

\renewcommand{\baselinestretch}{1.0}

\begin{figure*}[t]
\hspace*{-0.5cm}\psfig{figure=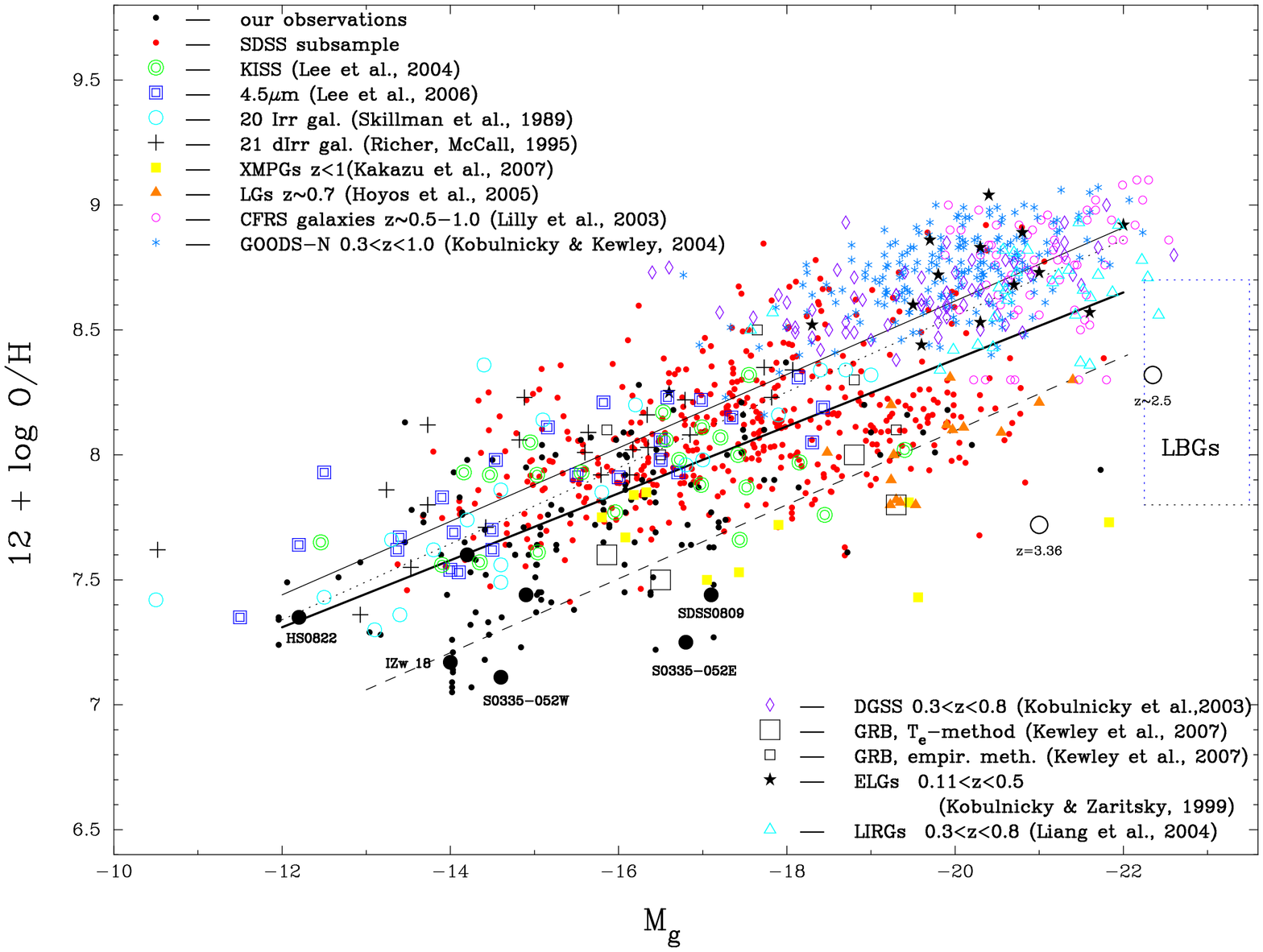,angle=-90,width=17.5cm,clip=}
\caption{Oxygen abundance vs absolute magnitude for the same galaxies as in 
Fig.~\ref{fig5} but all our galaxies including the SDSS sample
are shown by small filled circles. Additionally,
some well known  metal-poor galaxies, intermediate- and high-redshift 
galaxies are shown by symbols as labelled in the figure.
The position of the Lyman break galaxies (LBGs) by \citet{Pettini2001}
is indicated by the dotted line rectangle.
The thick solid line is a mean least-squares fit to all our data. The thin 
solid line is a least-squares fit to the data by \citet{RicherMcC1995} 
while the dotted  line is a mean least-squares fit to the data by  
\citet{Skillman1989}. The dashed line is the luminosity-metallicity relation 
for local metal-poor BCDs obtained by \citet{KunthOstlin2000}.}
\label{fig9}
\end{figure*}

\renewcommand{\baselinestretch}{1.5}

\section{Physical conditions and element abundances}

The electron temperature $T_{\rm e}$, the 
ionic and total heavy element abundances were derived 
following \citet{Iz06}. In particular, for the ions
O$^{2+}$, Ne$^{2+}$ and Ar$^{3+}$ we adopt
the temperature $T_{\rm e}$(O {\sc iii}) directly derived from the 
[O {\sc iii}] $\lambda$4363/($\lambda$4959 + $\lambda$5007)
emission-line ratio. 
For $T_{\rm e}$(O {\sc ii}) and $T_{\rm e}$(S {\sc iii}) we use the 
relation between the electron
temperatures $T_{\rm e}$(O {\sc iii}) and 
the temperatures characteristic for ions O$^{+}$ and S$^{2+}$  
obtained 
by \citet{Iz06}  from the H {\sc ii} photoionization models based on recent
stellar atmosphere models and improved atomic data \citep{Stasin2003}.

We use $T_{\rm e}$(O {\sc ii}) for the calculation of
O$^{+}$,  N$^{+}$, S$^{+}$ and Fe$^{2+}$ abundances and $T_{\rm e}$(S {\sc iii})
for the calculation of S$^{2+}$, Cl$^{2+}$ and Ar$^{2+}$ abundances.
The electron number densities for some H {\sc ii} regions 
were obtained from the [S {\sc ii}] $\lambda$6717/$\lambda$6731 emission line 
ratio. These lines were not observed or not measured in the remaining 
H {\sc ii} regions. 
For the abundance determination in those  H {\sc ii} regions
we adopt $N_{\rm e}$ = 10 cm$^{-3}$. The precise value of the electron number density
makes little difference in the derived abundances
since in the low-density limit which holds for the H {\sc ii} regions
considered here, the element abundances do not depend sensitively 
on $N_{\rm e}$.
The electron temperatures $T_{\rm e}$(O {\sc iii}), 
$T_{\rm e}$(O {\sc ii}), the ionization correction factors ($ICF$s), 
the ionic and total O and Ne abundances are given in
Table \ref{t5_1_36} for 3.6m observations.
  The electron temperatures
$T_{\rm e}$(O {\sc iii}), $T_{\rm e}$(O {\sc ii}), $T_{\rm e}$(S {\sc iii}), 
electron number density $N_{\rm e}$([S {\sc ii}]),
the ionization correction factors ($ICF$s), 
the ionic and total O, N, Ne, S, Cl, Ar and Fe abundances are given in 
Table \ref{t6_1_VLT} for VLT observations.

  The oxygen abundances 12 + log O/H in 61 H {\sc ii} regions out of 66 
obtained from the new 3.6m ESO and VLT observations range from 
7.05 to 8.22. Among them, 27 H {\sc ii} regions with 
12 + log O/H $<$ 7.6 are found, including 10 H {\sc ii} regions with 
12 + log O/H $<$ 7.3.
  The combined sample consisting of the new observations, 43 BCDs from the 
HeBCD sample, 30 galaxies from our previous 3.6m ESO observations and 20 
galaxies from the MMT observations yields a data set of 154 H {\sc ii} regions.
For comparison, we also use $\sim$9000
SDSS emission-line galaxies with the [O {\sc iii}] $\lambda$4363 emission 
line detected at least at the 1$\sigma$ level, allowing abundance 
determination by the direct $T_{\rm e}$-method. 
In addition, SDSS galaxies with high-quality spectra where 
the [O {\sc iii}]$\lambda$4363 emission line was not detected 
are used. 
In the latter case, the oxygen abundances were 
derived by the semiempirical method.
SDSS galaxies from the comparison sample
mostly populate the high-metallicity, high-luminosity ranges, 
as compared to the galaxies from our combined sample of low-metallicity
emission-line galaxies (Figs.~\ref{fig5} - \ref{fig7}). 
The considered galaxies span two dex in gas-phase oxygen abundance, from 
12 + log O/H $\sim$ 7.0 through $\sim$ 9.0.

We use SDSS $g$ magnitudes for the determination of the absolute 
magnitude $M_g$ of all galaxies from our samples, 
while usually $B$ magnitudes and $M_B$ are considered in the literature. 
However, \citet{Pap2008} have shown that for
regions with ongoing bursts of star formation, which 
is the case for our sample galaxies, the 
$B$--$g$ colour index is of the order of 0.1 mag only 
and $<$0.3 mag during the first few Gyrs of galactic evolution.
Therefore, we do not transform $M_g$ to $M_B$ and directly 
compare $M_g$'s for the galaxies from our samples with 
$M_B$'s for the galaxies available from the literature. 
The use of the SDSS $g$-band photometry for all our samples allows us to 
investigate the $L-Z$ relation over the $M_g$ range from --21 mag to the 
faintest magnitude of $\sim$ --12 mag at the low-metallicity end.

\section{Results}

\subsection{Luminosity-metallicity relation}

In order to illustrate the main properties of our sample we plot 
(a) the reddening parameter $C$(H$\beta$) obtained from the
Balmer decrement and (b) the logarithm of the H$\beta$ 
equivalent width (Fig.~\ref{fig3}) and the logarithm 
of the H$\beta$ line luminosity (in erg s$^{-1}$)
(Fig.~\ref{fig4}) as a function of absolute magnitude $M_g$. 
The new 3.6m telescope and VLT data are shown by open circles and stars, 
respectively.
The metal-poor galaxies collected from previous 3.6m ESO observations 
are shown by filled triangles \citep{BJlarge2007,Pap2008}.
Filled circles denote the data from the HeBCD sample collected 
by \citet{ING2004a} and \citet{IT04a}. 
The MMT data \citep{IT2007} are shown by large filled circles.
The comparison SDSS sample is represented by asterisks. 
From the latter sample H {\sc ii} regions in 
nearby spiral galaxies are excluded, as are 
faint SDSS galaxies with $m_g$ $>$ 18, the nearest 
SDSS galaxies with the redshift $z$ $<$ 0.004 and all SDSS galaxies 
with $\sigma$[$I(4363)$]/$I(4363)$ $>$ 0.25, totaling 443 SDSS galaxies 
from the comparison sample.

Our sample does not show any trend with absolute magnitude of either 
$C$(H$\beta$) or EW(H$\beta$), contrary to what was obtained by  
\citet{Salzer05} for the KISS sample.
The extinction in our sample galaxies is low. Only a few galaxies 
have $C$(H$\beta$) $>$ 0.4. 
The range of EW(H$\beta$) $\sim$ 0 -- 300 $\AA$ for the galaxies from
our sample is similar to that for the KISS sample  
\citep{Salzer05} but it is higher than that for the high-redshift galaxies 
of \citet{Kobulniky2003} where EW(H$\beta$) $\leq$ 60 $\AA$.
 
The logarithm of the H$\beta$ luminosity log $L$(${\rm{H}\beta}$) of our
galaxies ranges from 36 to 42 (Fig.~\ref{fig4}). 
For comparison, the galaxies from the KISS sample by \citet{Salzer05} and
intermediate-redshift galaxies by \citet{Kobulniky2003} have
log $L$(${\rm{H}\beta}$) $\sim$ 39 -- 43 and 39 -- 42, respectively,
i.e. low-luminosity galaxies are lacking.
 
In Fig.~\ref{fig5} we show the oxygen abundance - absolute magnitude $M_g$
relation for the galaxies with oxygen abundances calculated 
mainly with the $T_{\rm e}$-method. 
In this Figure, the same samples and symbols as in Fig~\ref{fig3} are used.
The region denoted as ``branch'' is populated mainly by galaxies
with relatively high redshifts ($z$ $>$ 0.02) and oxygen abundances derived by 
the $T_{\rm e}$-method.
Note that selection effects could be present for ``branch'' high-redshift 
galaxies which are predominantly distant spirals. In these galaxies we select mainly 
low metallicity \h2\ regions 
with a detectable [O {\sc iii}]$\lambda$4363 line while the abundance 
gradient is present in spirals. 
The dotted line is a mean least-squares fit to all our data and the solid line 
is a mean least-squares fit to our data excluding ``branch'' galaxies 
with M$_g$ $<$ --18.4 and systems with an oxygen abundance 
in the range 8.0 -- 8.3.
The dashed line is a mean least-squares fit to the local dwarf 
irregular galaxies by \citet{Skillman1989}.
Our sample (including the SDSS subsample) shows the familiar trend of 
increasing metallicity with increasing luminosity. 
A linear least square fit to all data yields the relation

\begin{equation}
{\rm 12+log(O/H)} = (5.706\pm 0.199) - (0.134\pm 0.012) {\rm M}_{g}
\end{equation}
(dotted line in Fig.~\ref{fig5}).
Excluding ''branch'' galaxies we obtain the relation
\begin{equation}
{\rm 12+log(O/H)} = (5.076\pm 0.320) - (0.174\pm 0.200) {\rm M}_{g}
\end{equation}
(solid line in Fig.~\ref{fig5}).
We note that the Skillman et al. and Richer \& McCall 
fits do not extend over the metallicity range of the present data.
Therefore, we extrapolate the former fit in Fig.~\ref{fig5} (dashed line) 
to higher metallicities. 
Skillman's and our fits are obviosuly very similar. 
The slopes of our $L-Z$ relation of 0.134 (0.174) are very close to 
the slope of 0.153 by \citet{Skillman1989} and to the slope of 
0.147 by \citet{RicherMcC1995}.

Our sample is well populated in the low-luminosity range, while less than
10 galaxies from the KISS sample \citep{Salzer05} 
which were used for the study of the $L-Z$ relation 
are fainter than $M_B$ = --15, and 
none of them has an oxygen abundance less than 7.6.
Our sample, excluding the SDSS subsample, has a lower dispersion
around the dotted line compared to all our data and shows a shift to lower
metallicities or/and higher luminosities. 
This likely can be attributed to our selection criteria which 
are optimized for the search for very metal-poor emission-line galaxies.
Additionally, our sample galaxies display significant to strong 
ongoing SF giving rise to a large contribution from  
young stars and ionized gas to the total light of the galaxy. 
\citet[][see also \citet{P02}]{P96}, using surface 
brightness profile decomposition to separate the
star-forming component from the underlying host galaxy of BCDs,
found that SF regions provide on average 50\% of the total 
$B$-band emission within the 25 $B$ \sbb\ isophote, 
with several examples of more intense 
starbursts whose flux contribution exceeds 70\%. 
As a result, a shift of BCDs by a $\Delta\,M \sim$--0.75 mag with respect 
to the relatively quiescent dIrr population is to be expected in  
Fig.~\ref{fig5} (see also Fig.~\ref{fig9}).
A similar offset to lower metallicities or/and higher luminosities has been
found by \citet{Kakazu07} for their intermediate-redshift low-metallicity
emission-line galaxies with strong SF activity. 
The mass estimate of the galaxy is less sensitive to the presence 
of star-forming regions as compared to its luminosity. 
This was demonstrated by \citet{Ellison2008} who found 
that galaxies in close pairs show enhanced SF activity 
as compared to a control sample of isolated galaxies. 
At the same time galaxies in close pairs show a smaller offset in the 
mass-metallicity relation as compared  to the luminosity-metallicity relation.
Thus, the offset in Fig.~\ref{fig5} indicates that both higher luminosities and
lower metallicities may contribute to the shift in the luminosity-metallicity
diagram of our sample galaxies relative to more quiescent dIrrs.

In Fig.~\ref{fig6} we demonstrate that the region of ``branch'' 
galaxies is populated mainly by relatively high-redshift systems.
The sample is the same as in Fig.~\ref{fig5}
but in the left panel only SDSS galaxies with oxygen abundances derived with 
the $T_{\rm e}$-method are shown and in the right panel only relatively 
high-redshift galaxies with $z$ $>$ 0.02 are selected.

The location of the galaxies on the luminosity-metallicity diagram is also
sensitive to the method used for the abundance determination.
In order to illustrate its effect on the observed $L-Z$ relation,
we compare in Fig.~\ref{fig7} the oxygen abundance of SDSS sample galaxies
(dots) obtained with the direct $T_{\rm e}$-method (left panel) and with the 
semiempirical strong-line method (right panel).
The abundances for other galaxies in Fig.~\ref{fig7} are derived with the 
$T_{\rm e}$-method.
In this Figure we show the larger control sample of the SDSS ($N$=7964) 
as compared to Fig.~\ref{fig5}. 
Only H {\sc ii} regions in nearby spiral galaxies and from the 
nearest SDSS galaxies with redshifts $z$ $<$ 0.004 were excluded 
from the $\sim$ 9000 SDSS sources while faint galaxies with $m_g$ $>$ 18 
are included. 
Symbols in Fig.~\ref{fig7} are the same as in Fig.~\ref{fig3} except for SDSS 
galaxies which are shown by dots. 
The dotted line is a mean least-squares fit to all our data from 
Fig.~\ref{fig5}, while the solid line is a mean 
least-squares fit to the same data excluding ``branch'' galaxies.

It can be seen from Fig.~\ref{fig7} that the oxygen abundance of a given
galaxy obtained by different methods could differ by $\sim$0.3--0.5 dex,  
especially for luminous galaxies. 
This figure illustrates clearly above 
12 + log O/H $\sim$ 8.5 and $M_g$ $<$ --19 - --20
significant discrepancies between oxygen abundances 
obtained from the $T_{\rm e}$-method and empirical methods.
\citet{Stas2002} emphasized that, at high metallicity, the $T_{\rm e}$
derived from [O {\sc iii}] $\lambda$4363 would largely overestimate
the temperature of the O$^{++}$ zone (and largely underestimate the metallicity)
because cooling is dominated by the [O {\sc iii}] $\lambda$52 $\mu$m and 
[O {\sc iii}] $\lambda$88 $\mu$m lines.
At the same time \citet{Pil2007} demonstrated that there is
an observational limit of the highest possible metallicities 
near 12 + log O/H $\sim$ 8.95. 
This maximum value was determined in the centers of the most luminous 
(--22.3 $\la$ $M_B$ $\la$ --20.3) galaxies using the 
semiempirical ff-method \citep{Pil2006}. 
Thus, although the main mechanisms determining the electron 
temperature in H {\sc ii} nebulae have been known for a long time, 
there are still important unsolved problems. 

The contribution of star-forming regions to the light of the galaxy can be
quantified by the equivalent width EW(H$\beta$) of the H$\beta$ emission line
which in turn depends on the age of the burst of star formation.
In Fig.~\ref{fig8} we show the same samples as in Fig.~\ref{fig5} except for
the SDSS galaxies now being split into two subsamples. In the 
left panel only those with high equivalent widths 
EW(H$\beta$) $>$ 80$\AA$ are shown while in the right panel only SDSS 
galaxies with low equivalent widths EW(H$\beta$) $<$ 20$\AA$ are plotted.
The dotted line in the left and right panels is a 
mean least-squares fit to all our data shown in Fig.~\ref{fig5}, 
while the solid line is a mean 
least-squares fit to the same data excluding ``branch'' galaxies.
There is a clear difference between the two subsamples of the SDSS galaxies 
by $\sim$ 0.4 dex in oxygen abundance
or, equivalently, by $\sim$ 3 mag in absolute magnitude. 
SDSS galaxies with EW(H$\beta$) $>$ 80$\AA$ nicely follow the relation
for our dwarf low-metallicity emission-line galaxies shown as reference objects
by filled and open circles, stars, filled triangles and large filled circles. 
On the other hand,
the SDSS galaxies with EW(H$\beta$) $<$ 20$\AA$ are located systematically
above the low-metallicity galaxies. We propose two possible
explanations for such a difference between the two subsamples of 
SDSS galaxies: 1) the emission of the SDSS galaxies with high EW(H$\beta$)
is dominated by star-forming regions, therefore they have higher
luminosities compared to galaxies in a relatively 
quiescent stage; 2) SDSS galaxies
with low EW(H$\beta$) are the ones with higher astration level, therefore they
are more chemically evolved systems with higher oxygen abundances.
Perhaps both of these explanations are tenable, accounting for the 
observed differences between SDSS galaxies with high and low EW(H$\beta$).
Thus, the lowest-metallicity SDSS galaxies are found predominantly among
galaxies with 
high EW(H$\beta$). 
On the other hand, no extremely low-metallicity SDSS 
galaxies are found among systems with EW(H$\beta$) $<$ 20$\AA$.
Thus, mixing of the SDSS galaxies with EW(H$\beta$) $<$ 20$\AA$ and 
$>$ 80$\AA$ results in a significant increase of the dispersion of the 
luminosity-metallicity diagram. 

The redshift of the galaxy could also play a role. In Fig.~\ref{fig5} the
  bulk of the galaxies with 12+log(O/H)=8.0 -- 8.3 and absolute
  magnitudes between --19 and --21 mag (denoted as ``branch'' 
galaxies) is represented by higher-redshift
  systems as compared to other galaxies from the SDSS 
and a correction for redshift is required. 
Since ``branch'' galaxies are blue, a correction for redshift 
for systems with weak emission lines would increase 
their brightness by $\sim$ 0.1 - 0.3 mag. This will not be 
enough to remove the offset between ``branch'' galaxies 
and lower-redshift galaxies in Fig.~\ref{fig5}.
The situation is more complicated for ``branch'' galaxies with strong
emission lines since their effect on the apparent magnitudes of a galaxy 
in standard passbands will significantly depend on redshift 
\citep[see e.g.][]{Z08}.
Because of these reasons, we 
decided not to take into account corrections for redshift.

\subsection{Comparison of our sample with other data}

In Fig.~\ref{fig9} we compare our $L-Z$ relation with other 
published data for galaxies of different types.
In this Figure, all of our galaxies from Fig.~\ref{fig5}, including
those from the comparison SDSS sample, are shown by small filled circles.
Some well known  metal-poor galaxies are depicted by large filled circles and 
are labelled. 
Their absolute magnitudes $M_B$ are taken from \citet{Kewley2007}. 
For comparison, 23 KISS emission-line galaxies by
\citet{Lee2004} are displayed with large open double circles. The abundances 
for these galaxies are derived with the $T_{\rm e}$-method, 
the $B$-band magnitudes are from \citet{Salzer1989} and \citet{GildePaz2003}.
With open double squares we show 25 nearby dIrrs with the 4.5$\mu$m 
{\sl Spitzer} luminosities and compiled O/H abundances derived with
the $T_{\rm e}$-method \citep{Lee45mu2006}. With large open circles and large 
crosses we respectively show
20 irregular galaxies from \citet{Skillman1989} and 21 dwarf irregular 
galaxies from \citet{RicherMcC1995} for which oxygen abundances
are obtained mainly with the $R_{23}$ empirical method, and for a few
  objects only with the $T_{\rm e}$-method.
The thick solid line is a 
mean least-squares fit to all our data. The thin solid line is a 
least-squares fit to the data by \citet{RicherMcC1995} 
while the dotted  line is a mean least-squares fit to the data by  
\citet{Skillman1989}. 
The dashed line is the luminosity-metallicity relation  for local 
metal-poor BCDs obtained by \citet{KunthOstlin2000}.

Data for intermediate- and high-redshift galaxies are also shown. 
The most distant ($z$ $<$ 1) extremely metal-poor galaxies (XMPGs) 
 \citep{Kakazu07} with the oxygen abundances derived with the empirical method 
are shown with filled squares, while relatively metal-poor luminous galaxies 
at $z$ $\sim$ 0.7 \citep{Hoyos2005} (O/H derived with the $T_{\rm e}$-method)
with filled triangles. 
The remaining samples in Fig. \ref{fig9} are the following:
a) the large open circles correspond to the $z$ = 3.36 lensed galaxy 
\citep{Villar2004} and to the average position of luminous Lyman-break 
galaxies at redshifts $z$ $\sim$ 2.5 \citep{KobKoo2000} (O/H derived
with the $R_{23}$ method); 
b) small open circles stand for 66 Canada-France 
Redshift Survey (CFRS) galaxies by \citet{Lilly2003} in the redshift range of
$\sim$ 0.5 -- 1.0 (O/H derived with the $R_{23}$ empirical method);
c) asterisks are for 204 GOODS-N (Great Observatories Origins Deep Survey - 
North) emission-line galaxies in the range of redshifts 0.3 $<$ $z$ $<$ 1.0
\citep{KobKew04} (O/H is derived with the $R_{23}$ empirical method);
d) small open rombs indicate 64 emission-line field galaxies from the 
Deep Extragalactic Evolutionary 
Probe Groth Strip Survey (DGSS) in the redshift range of $\sim$ 0.3 -- 0.8
\citep{Kobulniky2003} (O/H derived with the $R_{23}$ empirical method);
e) open squares are for the gamma-ray burst (GRB) hosts by \citet{Kewley2007}.
Small open squares are for galaxies with O/H derived with the empirical method
\citep{KewleyDopita2002} and large open squares for the galaxies with
O/H derived with the $T_{\rm e}$-method \citep{Kewley2007}; 
f) filled stars denote the 14 star-forming emission-line galaxies at 
intermediate redshifts (0.11 $<$ $z$ $<$ 0.5) by \citet{KobylZarit1999} 
(O/H derived with the empirical $R_{23}$ method); 
g) open filled triangles are for 
 29 distant 15$\mu$m-selected luminous infrared galaxies (LIRGs) at 
$z$ $\sim$ 0.3 -- 0.8 taken from the sample of \citet{Liang2004}
(O/H derived with the empirical method); 
and, finally, h) the large dotted rectangle depicts the position
of Lyman break galaxies \citep[LBGs, ][]{Pettini2001} on the $L-Z$ diagram.

The location of our galaxies on the luminosity-metallicity diagram is
similar to that obtained previously for local emission-line galaxies
but is shifted to higher luminosities and/or lower metallicities compared 
to that obtained for quiescent irregular dwarf galaxies.
For comparison, \citet{Lee2004} have also 
demonstrated that their 54 H {\sc ii} KISS
galaxies with O/H derived with the $T_{\rm e}$-method follow the $L-Z$ 
relation with a slope similar to that for a more quiescent dIrrs 
but are shifted to higher brightness by 0.8 magnitudes.
Furthermore, they have shown that H {\sc ii} galaxies with disturbed 
irregular outer 
isophotes (likely due to the interaction) are shifted to a more luminous 
and/or more metal-poor region in the $L-Z$ diagram as compared to  
morphologically more regular galaxies. Note that their samples of 
H {\sc ii} galaxies and of dIrrs are in the same luminosity range 
as our sample.
  \citet{Pap2008} also note that in contrast to the majority ($>$90\%) of 
BCDs, the extremely metal-poor SF dwarfs reveal more irregular and bluer hosts.

Thus, the difference in the zero point between our $L-Z$
relation for low-metallicity galaxies and for other galaxies seems 
to be primarily due to the differences in the intrinsic properties of the
galaxies selected for different samples with various selection criteria.

A key question is whether 
a unique $L-Z$ relation does exist for galaxies of different types. 
The assessment of this issue is complicated by offsets of 
high-redshift galaxies with different look-back-times. 
In this context, \citet{Kobulniky2003} 
have shown that both the slopes and zero points of the 
$L-Z$ relation exhibit a smooth evolution with redshift. 
A possible universal $L-Z$ relation for galaxies is also blurred 
by the fact that metallicity determinations of various galaxy samples,
differing in their EW(H$\beta$), absolute magnitude and redshift, do not employ
a unique technique. More specifically, 
several authors emphasize the presence of a well-known shift between 
the O/H ratio obtained by the direct $T_{\rm e}$-method and empirical 
strong-line methods. 
Oxygen abundances obtained by empirical methods are by 0.1 --0.25 dex 
\citep{Shi2005} and even by up to 0.6 dex \citep{Hoyos2005} higher than
those obtained with the $T_{\rm e}$-method. 
For our sample we obtained an offset of $\sim$0.3--0.5 dex.

It can be seen from Fig.~\ref{fig9} that the high-redshift galaxies 
with an oxygen abundance derived by the $T_{\rm e}$-method have 
a shallower slope compared to local galaxies. 
On the other hand, oxygen abundances of high-redshift galaxies 
obtained with the $R_{23}$ empirical strong-line method 
\citep[data in Fig. \ref{fig9} by ][]{Lilly2003,KobKew04,Liang2004} are
higher and follow the relation for high-metallicity SDSS galaxies in
Fig. \ref{fig7}b despite the fact that oxygen abundances for the latter 
galaxies were calculated with the different semi-empirical strong-line method.
Because of this agreement we decided not to re-calculate oxygen abundances
of high-redshift galaxies with the semi-empirical method and adopted O/H 
values from the literature.
Keeping in mind the systematic differences between oxygen abundances derived 
with the empirical and the $T_{\rm e}$-methods, it might be worth
considering a decrease in oxygen abundance by $\sim$ 0.2 -- 0.6 dex
for all high-redshift galaxies with O/H derived with the empirical method. 
In that case, the position of 
high-redshift galaxies on the $L-Z$ diagram would be consistent with that of
the ``branch'' galaxies.
Such considerations add further support to the  
results obtained by \citet{Pil2004} and \citet{Shi2005}
that the more luminous galaxies have a slope of the $L-Z$ 
relation more shallow than that of the dwarf galaxies.

We presume that our $L-Z$ relation could be useful as a local 
reference for studies of this relation for other types 
of local galaxies and/or of high-redshift galaxies.

\section{Summary}

We present VLT spectroscopic observations of a new 
sample of 28 H {\sc ii} regions from 16 emission-line galaxies 
and 
ESO 3.6m telescope spectroscopic observations 
of a new sample of 38 H {\sc ii} regions from 28 emission-line galaxies.
These galaxies have mainly been selected from the Data Release 6 (DR6) of the
Sloan Digital Sky Survey (SDSS) as low-metallicity galaxy candidates.

Physical conditions and element abundances are derived with the 
$T_{\rm e}$-method 
for 38 H {\sc ii} regions observed with the 3.6m telescope and for 
23 H {\sc ii} regions observed with the VLT.

  From our new observations we find  that the oxygen abundance in 
61 out of the 66 observed H {\sc ii} in our sample
ranges from 12 + log O/H = 7.05 to 8.22. 
The oxygen abundance in 27 H {\sc ii} regions is
12 + log O/H $<$ 7.6 and among them 10  H {\sc ii} regions have 
an oxygen abundance less than 7.3.

This new data in combination with objects from 
our previous studies constitute a large uniform sample 
of 154 H {\sc ii} regions with high-quality spectroscopic 
data which are used to study the luminosity-metallicity 
($L-Z$) relation for the local galaxies with emphasis on its
  low-metallicity end.

  As a comparison sample we use $\sim$ 9000 SDSS emission-line galaxies with
higher oxygen abundances which are also obtained mainly 
by the direct $T_{\rm e}$-method.
For all of our sample galaxies the $g$ magnitudes are taken from the SDSS
while the distances are from the NED.
The entire sample spans nearly two orders of magnitude 
with respect to its gas-phase metallicity, from 
12 + log O/H $\sim$ 7.0 to $\sim$ 8.8, and covers 
an absolute magnitude range from $M_g$ $\sim$ --12 
to $\sim$ --20. 

We find that the metallicity-luminosity relation for our galaxies 
is consistent with previous ones obtained for objects of similar 
type.
 The local $L-Z$ relation obtained with high-quality spectroscopic data 
 is useful for predictions of galaxy evolution models. 

\begin{acknowledgements}
N. G. G. and Y. I. I. thank the Max Planck Institute for Radioastronomy (MPIfR)
for hospitality, and acknowledge support through DFG grant 
No. Fr 325/57-1.
P. P. thanks the Department of Astronomy and Space Physics at 
Uppsala University for its warm hospitality.
K. J. Fricke thanks the MPIfR for Visiting Contracts during 2008 and 2009.
This research was partially funded by project grant 
AYA2007-67965-C03-02 of the Spanish Ministerio de Ciencia e Innovacion.
We acknowledge the work of the Sloan Digital Sky Survey (SDSS) team.
Funding for the SDSS has been provided by the Alfred P. Sloan Foundation, the
Participating Institutions, the National Aeronautics and Space Administration,
the National Science Foundation, the US Department of Energy, the Japanese
Monbukagakusho, and the Max Planck Society. The SDSS Web site is http://
www.sdss.org/.
\end{acknowledgements}

\Online

\renewcommand{\baselinestretch}{1.0}



\begin{table*}[tbh]
\caption{List of galaxies observed with the 3.6m ESO telescope}
\label{obs36}
\begin{tabular}{lcclcccc} \hline \hline

Galaxy & R.A., Dec.     &  Date &  Exp. time      &  Airmass   & seeing & redshift  \\ 
       &        (J2000) &(2007) &  (sec)          &            &        &        \\ \hline
 
J0015+0104       &~\ra{00}{15}{20}{7},~~~\dec{01}{04}{37}& 14 Sep  & 4$\,\times\,$900 & 1.207 & 1\farcs74 & 0.00689 $\pm$ 0.00021  \\
J0016+0108           &~~00~~16~~~28.~3,~~~01~~08~~02 & 15 Sep  & 4$\,\times\,$675   & 1.306 & 1\farcs90 & 0.01035 $\pm$ 0.00006  \\
J0029-0108           &~~00~~29~~~04.~7,~--01~~08~~26 & 16 Sep  & 4$\,\times\,$450  & 1.614 & 1\farcs88 & 0.01313 $\pm$ 0.00006  \\
J0029-0025           &~~00~~29~~~49.~5,~--00~~25~~40 & 14 Sep  & 4$\,\times\,$900  & 1.142 & 1\farcs72 & 0.01440 $\pm$ 0.00013  \\
0057-0022            &~~00~~57~~~12.~6,~--00~~21~~58 & 16 Sep  & 4$\,\times\,$900  & 1.525 & 1\farcs47 & 0.00956 $\pm$ 0.00023  \\
J0107+0001           &~~01~~07~~~50.~8,~~~00~~01~~28 & 16 Sep  & 3$\,\times\,$267  & 1.188 & 1\farcs41 & 0.01835 $\pm$ 0.00012  \\
\end{tabular}

\end{table*}


\begin{table*}[tbh]
\caption{List of galaxies observed with the VLT$^a$}
\label{obsVLT}
\begin{tabular}{lcclcccc} \hline \hline

Galaxy & R.A., Dec. &  Date &  Exp. time      &  Airmass   & seeing & redshift  \\ 
       &       (J2000) &       &  (sec)          &            &  &        \\ \hline

J0004+0025 No.1 &~\ra{00}{04}{21}{6},~~~\dec{00}{25}{36} & 20.10.2006& 2$\,\times\,$750& 1.246&0\farcs60 & 0.01269 $\pm$ 0.00004  \\                     
                      &                              &            &2$\,\times\,$540  & 1.181 & 0\farcs37 & 0.01269 $\pm$ 0.00004  \\
J0004+0025 No.2  &~~00~~04~~~21.~6,~~~00~~25~~36     & 20.10.2006 &2$\,\times\,$750  & 1.246 & 0\farcs60 & 0.01266 $\pm$ 0.00003  \\                    
                      &                              &            &2$\,\times\,$540  & 1.181 & 0\farcs37 & 0.01266 $\pm$ 0.00003  \\
J0014-0044 No.1  &~~00~~14~~~28.~8,~--00~~44~~44     & 19.11.2006 &2$\,\times\,$750  & 1.158 & 1\farcs22 & 0.01361 $\pm$ 0.00004  \\
                     &                               &            &2$\,\times\,$540  & 1.217 & 0\farcs93 & 0.01361 $\pm$ 0.00004  \\
\end{tabular}

$^a$first line for each galaxy is related to the observation in the
blue range and second line to the one in the red range.
\end{table*}

{\scriptsize
  \begin{longtable}{lrrrrrrrr}
  \caption{Emission line intensities (3.6m ESO observations)}
\label{t3_1_36} \\
\hline \hline
\endfirsthead
\caption{continued.}\\
\hline\hline
  {Ion}
  &{$I$($\lambda$)/$I$(H$\beta$)}
  &{$I$($\lambda$)/$I$(H$\beta$)}
  &{$I$($\lambda$)/$I$(H$\beta$)}
  &{$I$($\lambda$)/$I$(H$\beta$)}
  &{$I$($\lambda$)/$I$(H$\beta$)}
  &{$I$($\lambda$)/$I$(H$\beta$)}
  &{$I$($\lambda$)/$I$(H$\beta$)}
  &{$I$($\lambda$)/$I$(H$\beta$)} \\ \hline
\endhead
\hline
\endfoot
  {Ion}
  &{$I$($\lambda$)/$I$(H$\beta$)}
  &{$I$($\lambda$)/$I$(H$\beta$)}
  &{$I$($\lambda$)/$I$(H$\beta$)}
  &{$I$($\lambda$)/$I$(H$\beta$)}
  &{$I$($\lambda$)/$I$(H$\beta$)}
  &{$I$($\lambda$)/$I$(H$\beta$)}
  &{$I$($\lambda$)/$I$(H$\beta$)}
  &{$I$($\lambda$)/$I$(H$\beta$)} \\ 
\cline{2-9}
  {}&\multicolumn{8}{c}{\sc Galaxy} \\     \cline{2-9}
  {}&
 {J0015$+$0104       }&{J0016$+$0108       }&
 {J0029$-$0108       }&{J0029$-$0025       }&
 {J0057$-$0022       }&{J0107$+$0001       }&
 {J0109$+$0107       }&{J0126$-$0038       } \\ 
 & & & & & & & & No.1 \\ \hline
3727 [O {\sc ii}]                 & 118.7 $\pm$   5.7 & 221.2 $\pm$   7.8 & 171.6 $\pm$  30.8 & 130.3 $\pm$   7.8 & 219.7 $\pm$   5.4 & 162.3 $\pm$  13.6 & 209.3 $\pm$   4.3 & 238.1 $\pm$   4.1 \\
3868 [Ne {\sc iii}]               &   ...~~~~~~ &   ...~~~~~~ &   ...~~~~~~ &   ...~~~~~~ &  31.0 $\pm$   2.0 &    ...~~~~~~  &  30.1 $\pm$   1.0 &  38.5 $\pm$   0.9 \\
3889 He {\sc i} + H8              &    ...~~~~~~  &    ...~~~~~~  &    ...~~~~~~  &    ...~~~~~~  &  17.3 $\pm$   1.8 &    ...~~~~~~  &  20.6 $\pm$   1.6 &  17.2 $\pm$   0.9 \\
3968 [Ne {\sc iii}] + H7          &    ...~~~~~~  &    ...~~~~~~  &    ...~~~~~~  &    ...~~~~~~  &  16.1 $\pm$   1.9 &    ...~~~~~~  &  25.6 $\pm$   1.5 &  22.2 $\pm$   0.9 \\
\end{longtable}
\hspace{1.cm}$^a$ zero value is assumed if a negative value is derived.

\hspace{1.cm}$^b$ in units of 10$^{-16}$ erg s$^{-1}$ cm$^{-2}$.
}


{\scriptsize
  \begin{longtable}{lrrrrrrrr}
  \caption{Emission line intensities (VLT observations)}
\label{t4_1_VLT} \\
\hline \hline
\endfirsthead
\caption{continued.}\\
\hline\hline
  {Ion}
  &{$I$($\lambda$)/$I$(H$\beta$)}
  &{$I$($\lambda$)/$I$(H$\beta$)}
  &{$I$($\lambda$)/$I$(H$\beta$)}
  &{$I$($\lambda$)/$I$(H$\beta$)}
  &{$I$($\lambda$)/$I$(H$\beta$)}
  &{$I$($\lambda$)/$I$(H$\beta$)}
  &{$I$($\lambda$)/$I$(H$\beta$)}
  &{$I$($\lambda$)/$I$(H$\beta$)} \\ \hline
\endhead
\hline
\endfoot
  {Ion}
  &{$I$($\lambda$)/$I$(H$\beta$)}
  &{$I$($\lambda$)/$I$(H$\beta$)}
  &{$I$($\lambda$)/$I$(H$\beta$)}
  &{$I$($\lambda$)/$I$(H$\beta$)}
  &{$I$($\lambda$)/$I$(H$\beta$)}
  &{$I$($\lambda$)/$I$(H$\beta$)}
  &{$I$($\lambda$)/$I$(H$\beta$)}
  &{$I$($\lambda$)/$I$(H$\beta$)} \\
  \cline{2-9}
  {}&\multicolumn{8}{c}{\sc Galaxy} \\     \cline{2-9}
  {}&
 {J0004$+$0025   }&{J0004$+$0025   }&
 {J0014$-$0044   }&{J0014$-$0044   }&
 {J0202$-$0047   }&{J0301$-$0059   }&
 {J0301$-$0059   }&{J0301$-$0059   } \\
 & No.1 & No.2 & No.1 & No.2 & & No.1 & No.2 & No.3 \\ \hline
3727 [O {\sc ii}]                 & 214.05 $\pm$  11.27 &253.26 $\pm$  13.99 & 99.16 $\pm$   1.73 &298.77 $\pm$   9.30 &191.81 $\pm$   3.50 &331.24 $\pm$   6.29 &304.43 $\pm$   6.86 &330.51 $\pm$  21.26 \\
3750 H12                          &   ...~~~~~~         &  ...~~~~~~         &  4.57 $\pm$   0.56 &  ...~~~~~~         &  ...~~~~~~         &  ...~~~~~~         &  ...~~~~~~         &  ...~~~~~~         \\
3771 H11                          &   ...~~~~~~         &  ...~~~~~~         &  5.06 $\pm$   0.52 &  ...~~~~~~         &  4.22 $\pm$   0.66 &  ...~~~~~~         &  ...~~~~~~         &  ...~~~~~~         \\
3798 H10                          &   ...~~~~~~         &  ...~~~~~~         &  7.06 $\pm$   0.52 &  ...~~~~~~         &  4.83 $\pm$   0.67 &  ...~~~~~~         &  ...~~~~~~         &  ...~~~~~~         \\
\end{longtable}
$^a$ zero value is assumed if a negative value is derived.

$^b$ in units of 10$^{-16}$ erg s$^{-1}$ cm$^{-2}$.
}


{\scriptsize
  \begin{longtable}{lrrrrrrrr}
  \caption{Ionic and total heavy element abundances (3.6m ESO observations)}
\label{t5_1_36} \\
\hline \hline
\endfirsthead
\caption{continued.}\\
\hline\hline
\endhead
\hline
\endfoot
  &\multicolumn{8}{c}{\sc Galaxy} \\     \cline{2-9}
Property &
{J0015+0104        } & {J0016+0108        } &
{J0029-0108        } & {J0029-0025        } &
{J0057-0022        } & {J0107+0001        } &
{J0109+0107        } & {J0126-0038   } \\
 &  &  &  &  &  &  &  & No.1 \\ \hline
  $T_{\rm e}$(O {\sc iii}) (K) &
$20000  \pm1020 $ & $17023  \pm1008 $ &
$18580  \pm1188 $ & $19025  \pm1020 $ &
$16258  \pm 215 $ & $20000  \pm1054 $ &
$12337  \pm 558 $ & $14070  \pm 351 $  \\

  $T_{\rm e}$(O {\sc ii}) (K) &
$16284  \pm1293 $ & $15379  \pm1217 $ &
$15961  \pm1470 $ & $16083  \pm1273 $ &
$15006  \pm 180 $ & $16284  \pm1336 $ &
$12196  \pm 517 $ & $13624  \pm 313 $  \\

  \\
  O$^+$/H$^+$ ($\times$10$^4$) &
$ 0.084 \pm 0.017$ & $ 0.186 \pm 0.038$ &
$ 0.129 \pm 0.037$ & $ 0.096 \pm 0.019$ &
$ 0.199 \pm 0.008$ & $ 0.115 \pm 0.025$ &
$ 0.381 \pm 0.052$ & $ 0.294 \pm 0.020$ \\
  O$^{++}$/H$^+$ ($\times$10$^4$) &
$ 0.033 \pm 0.004$ & $ 0.153 \pm 0.022$ &
$ 0.096 \pm 0.019$ & $ 0.099 \pm 0.013$ &
$ 0.200 \pm 0.007$ & $ 0.053 \pm 0.007$ &
$ 0.571 \pm 0.076$ & $ 0.534 \pm 0.036$ \\
\end{longtable}
}


{\scriptsize
  \begin{longtable}{lrrrrrrrr}
  \caption{Ionic and Total Heavy Element Abundances (VLT observations)}
\label{t6_1_VLT} \\
\hline \hline
\endfirsthead
\caption{continued.}\\
\hline\hline
\endhead
\hline
\endfoot
  &\multicolumn{8}{c}{\sc Galaxy} \\     \cline{2-9}
Property&
{J0004$+$0025    } & {J0004$+$0025    } &
{J0014$-$0044    } & {J0014$-$0044    } &
{J0202$-$0047        } & {J0301$-$0059    } &
{J0301$-$0059    } & {J0301$-$0059    } \\
 &No.1  &No.2  &No.1  &No.2  &  &No.1  &No.2  &No.3 \\ \hline
  $T_{\rm e}$(O {\sc iii}) (K) &
$18630  \pm1019 $ & $16701  \pm1016 $ &
$13195  \pm 123 $ & $16308  \pm1006 $ &
$14607  \pm 277 $ & $12582  \pm 585 $ &
$13999  \pm1236 $ & $16248  \pm1033 $  \\

  $T_{\rm e}$(O {\sc ii}) (K) &
$15975  \pm1263 $ & $15229  \pm1221 $ &
$12939  \pm 112 $ & $15033  \pm1201 $ &
$14006  \pm 243 $ & $12416  \pm 539 $ &
$13571  \pm1104 $ & $15001  \pm1232 $  \\

  $T_{\rm e}$(S {\sc iii}) (K) &
$17163  \pm 846 $ & $15562  \pm 844 $ &
$13067  \pm 102 $ & $15236  \pm 835 $ &
$14307  \pm 230 $ & $12499  \pm 485 $ &
$13785  \pm1026 $ & $15186  \pm 857 $  \\

  $N_{\rm e}$(S {\sc ii}) (cm$^{-3}$) &
$  193 \pm 128$ & $  213 \pm 132$ &
$   33 \pm  28$ & $  167 \pm  78$ &
$   78 \pm  38$ & $   10 \pm  10$ &
$   10 \pm  10$ & $   10 \pm  10$  \\

  \\
\end{longtable}
}

\setcounter{figure}{0}
\begin{figure*}[t]
\hspace*{0.0cm}\psfig{figure=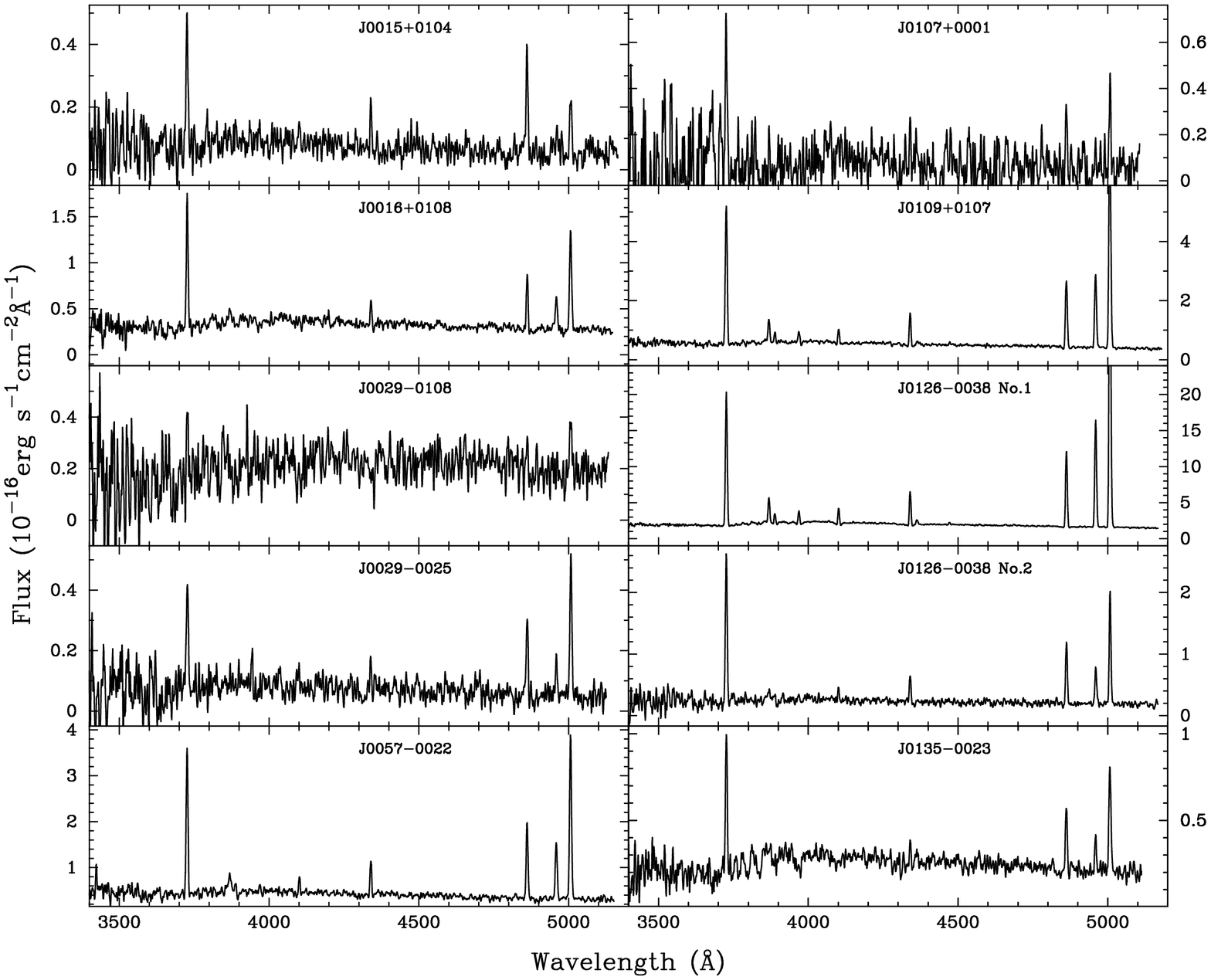,angle=-90,width=16.cm,clip=}
\caption{The flux-calibrated and redshift-corrected 3.6m ESO telescope spectra 
of the emission-line galaxies.
}
\label{fig1}
\end{figure*}

\setcounter{figure}{0}
\begin{figure*}[t]
\hspace*{0.0cm}\psfig{figure=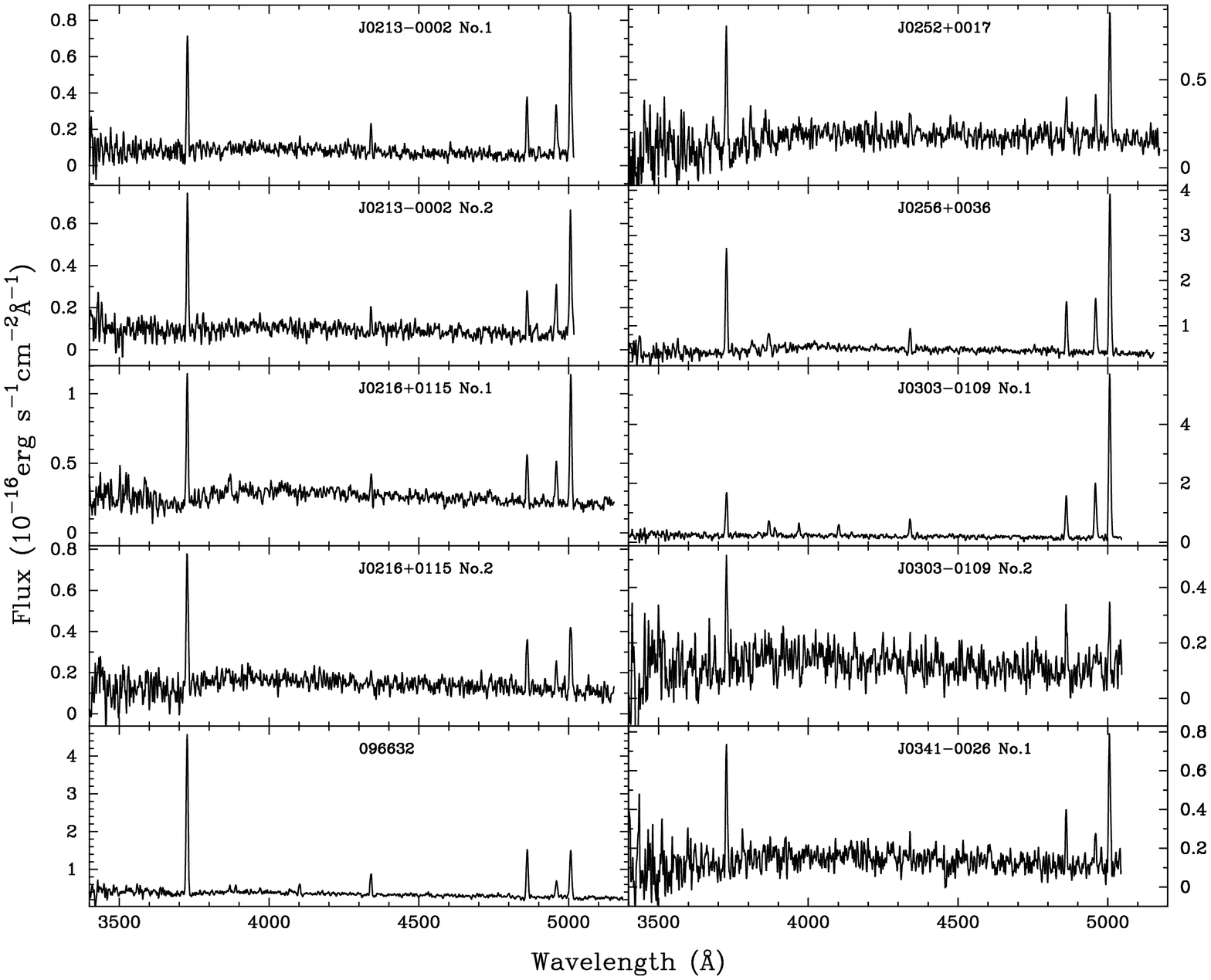,angle=-90,width=16.cm,clip=}
\caption{--$Continued$}
\end{figure*}

\setcounter{figure}{0}
\begin{figure*}[t]
\hspace*{0.0cm}\psfig{figure=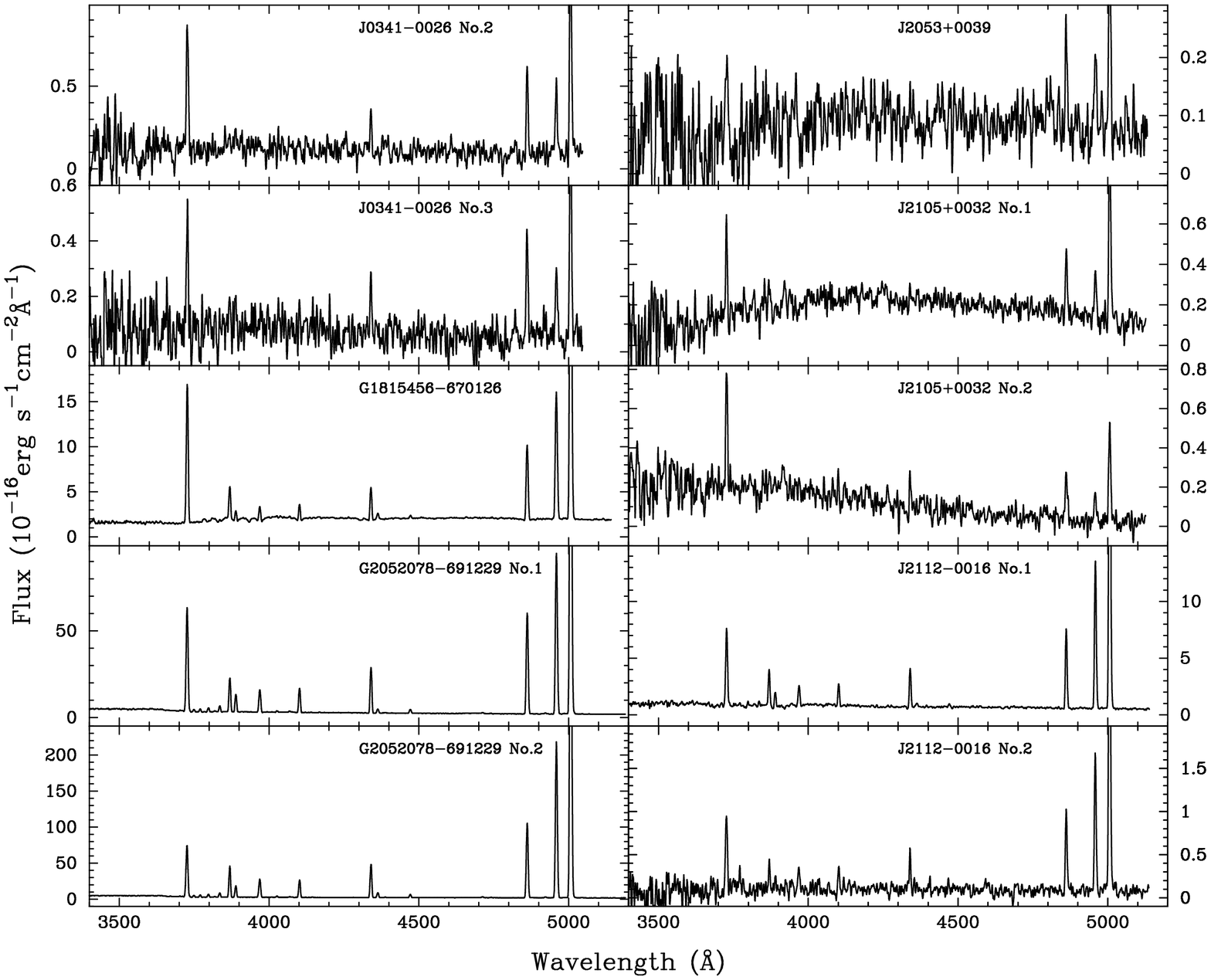,angle=-90,width=16.cm,clip=}
\caption{--$Continued$}
\end{figure*}

\setcounter{figure}{0}
\begin{figure*}[t]
\hspace*{0.0cm}\psfig{figure=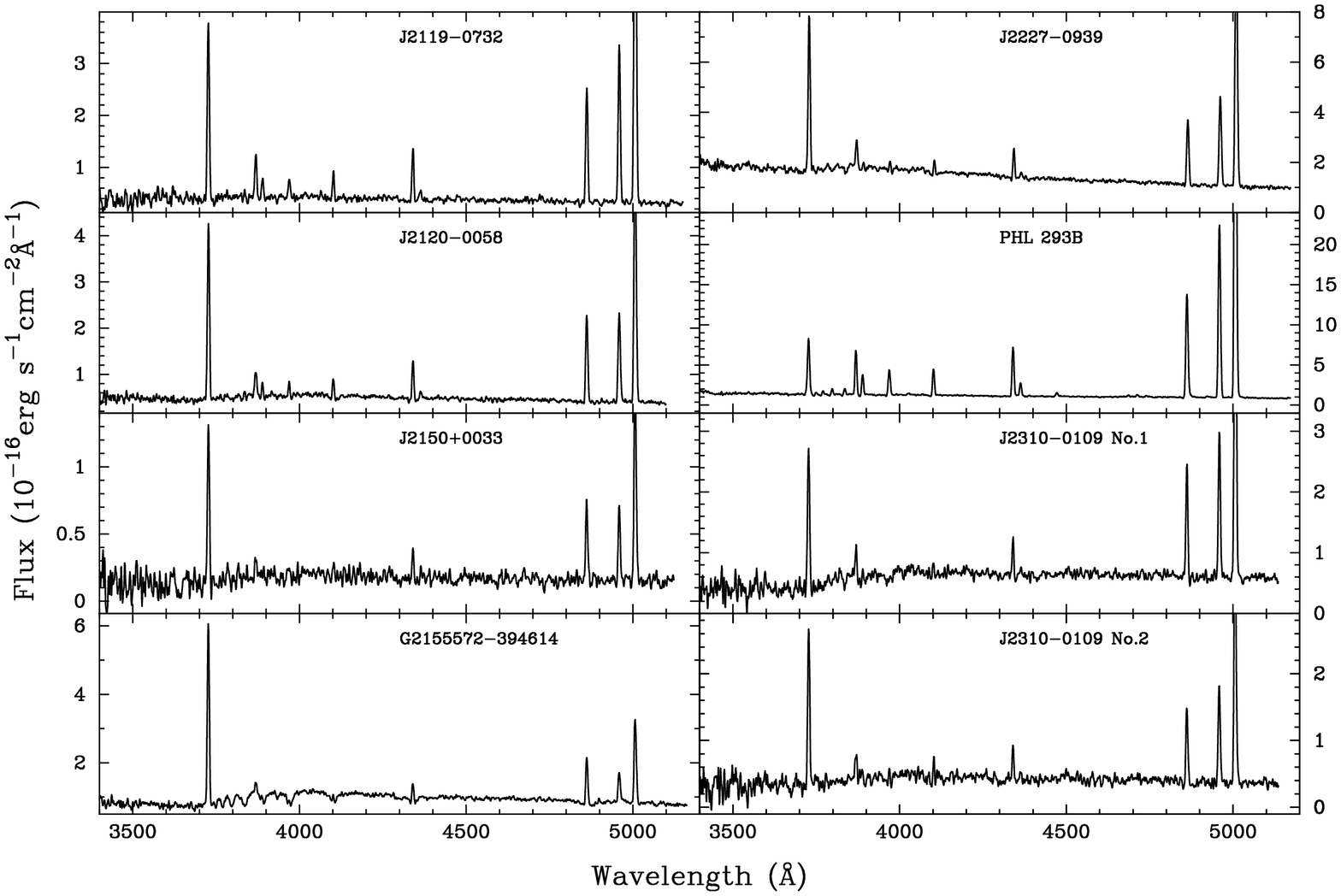,angle=-90,width=16.cm,clip=}
\caption{--$Continued$}
\end{figure*}

\setcounter{figure}{1}
\begin{figure*}[t]
\hspace*{0.0cm}\psfig{figure=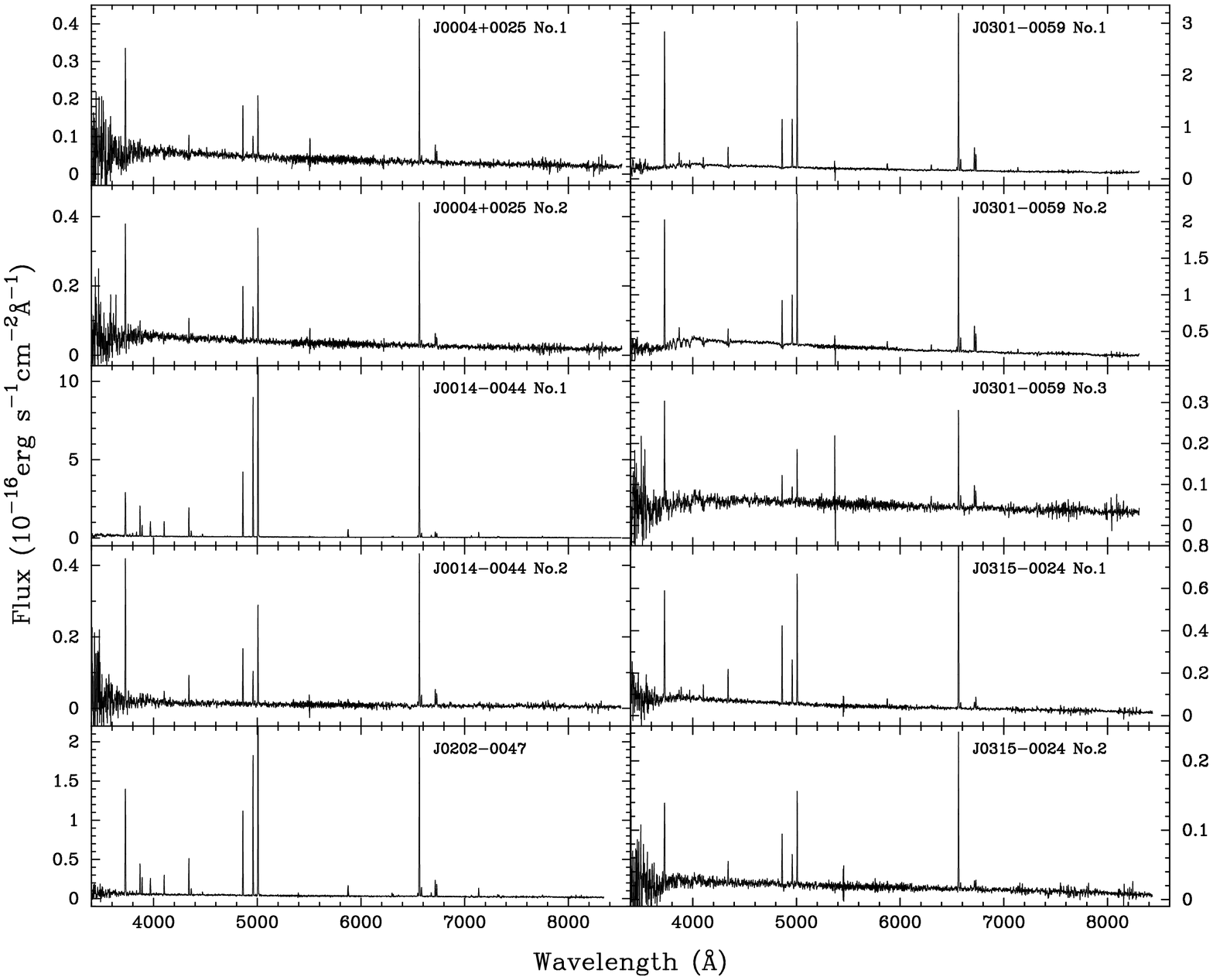,angle=-90,width=16.cm,clip=}
\caption{The flux-calibrated and redshift-corrected VLT spectra 
of the emission-line galaxies.
}
\label{fig2}
\end{figure*}

\setcounter{figure}{1}
\begin{figure*}[t]
\hspace*{0.0cm}\psfig{figure=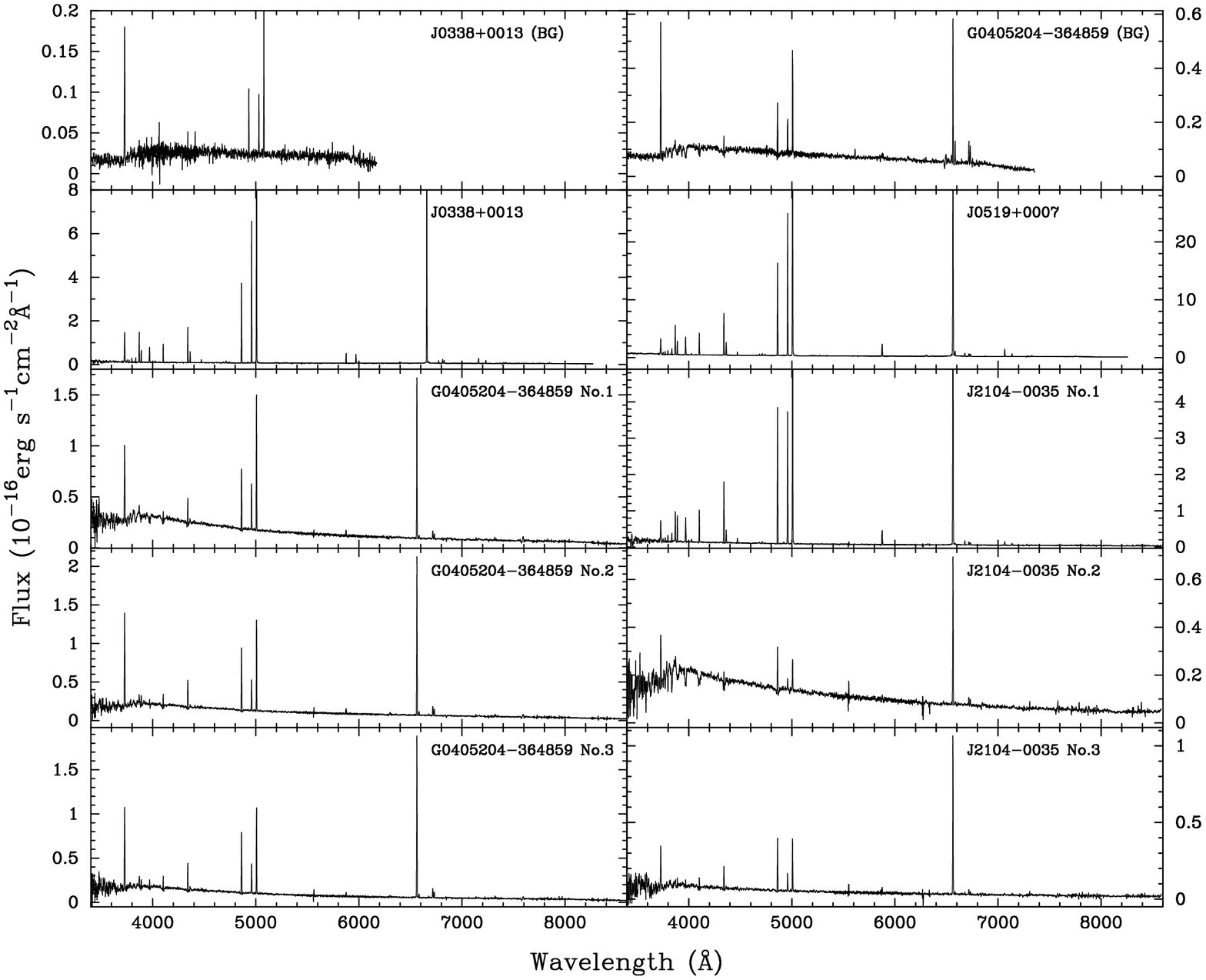,angle=-90,width=16.cm,clip=}
\caption{--$Continued$}
\end{figure*}

\setcounter{figure}{1}
\begin{figure*}[t]
\hspace*{0.0cm}\psfig{figure=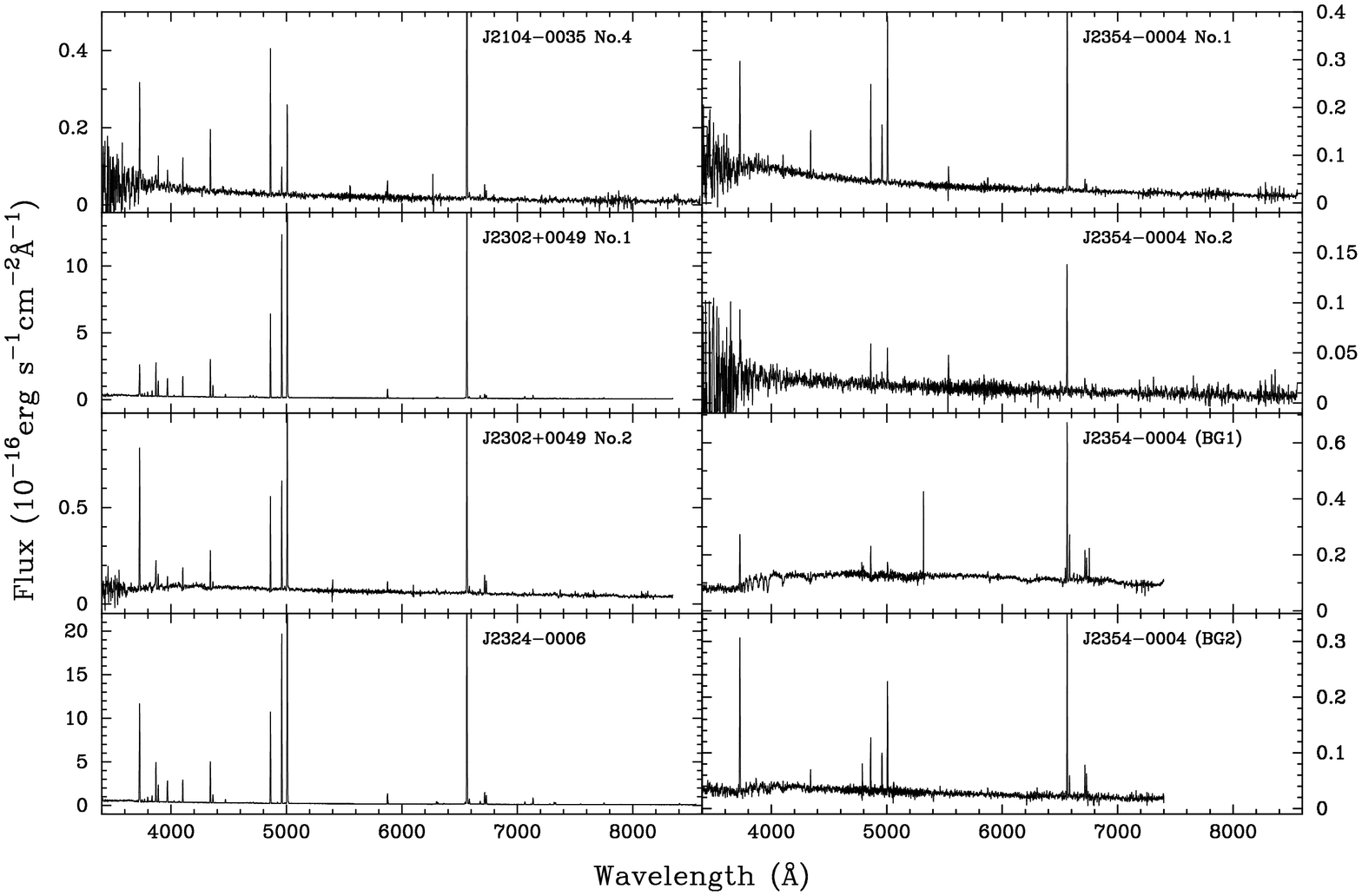,angle=-90,width=16.cm,clip=}
\caption{--$Continued$}
\end{figure*}


\begin{thebibliography}{}

\bibitem[Abazajian et al. (2009)]{A09} Abazajian, K., et al. 2009, \apjs,
182, 543

\bibitem[Contini et al. (2002)]{Contini2002} Contini, T., Treyer, M. A., Sullivan, M., 
\& Ellis, R. S. 2002, MNRAS, 330, 75 

\bibitem[Ellison et al. (2008)]{Ellison2008} Ellison, S. L., Patton, D. R., 
Simard, L., \& McConnachie, A. W. 2008, \aj, 135, 1877

\bibitem[Gavil\'an et al. (2009)]{Gavilan2009} Gavil\'an, M., Moll\'a, M., 
\& D\'iaz, \'A. 2009, to appear in {\sl Star-forming Dwarf Galaxies: Following
Ariadne’s Thread in the Cosmic Labyrinth}; preprint: arXiv0902.4695

\bibitem[Gil de Paz et al. (2003)]{GildePaz2003} Gil de Paz, A., 
Madore, B. F., \& Pevunova, O. 2003, \apjs, 147, 29

\bibitem[Guseva et al. (2007)]{BJlarge2007} Guseva, N. G., Izotov, Y. I.,
Papaderos, P., \& Fricke, K. J. 2007, \aap, 464, 885

\bibitem[Hoyos et al. (2005)]{Hoyos2005} Hoyos, C., Koo, D. C., Phillips, A. C., 
Willmer, C. N. A., \& Guhathakurta, P. 2005, \apj, 635, L21
 
\bibitem[Izotov et al. (1997)]{I97b} Izotov, Y. I., Lipovetsky, V. A.,
Chaffee, F. H., Foltz, C. B., Guseva, N. G., \& Kniazev, A. Y.
1997, \apj, 476, 698

\bibitem[Izotov \& Thuan (2004)]{IT04a} Izotov, Y. I., \& Thuan, T. X.
2004, \apj, 602, 200

\bibitem[Izotov \& Thuan (2007)] {IT2007} Izotov, Y. I. \& Thuan, T. X. 
2007, \apj, 665, 1115

\bibitem[Izotov et al. (2004a)]{ING2004a} Izotov, Y. I., Noeske, K. G.,
Guseva, N. G., Papaderos, P., Thuan, T. X. \& Fricke, K. J. 2004a, 
\aap, 415, L27

\bibitem[Izotov et al. (2004b)]{ISGT2004} Izotov, Y. I., Stasi\'nska, G., 
Guseva, N. G., \& Thuan, T. X. 2004b, \aap, 415, 87

\bibitem[Izotov et al. (2006a)]{Iz06} Izotov, Y. I., Stasi\'nska, G.,
Meynet, G., Guseva, N. G., \& Thuan, T. X. 2006a, \aap, 448, 955

\bibitem[Izotov et al. (2006b)]{IPGFT2006}  Izotov, Y. I., Papaderos, P., 
Guseva, N. G., Fricke, K. J., \& Thuan, T. X. 2006b, \aap, 454, 137 

\bibitem[Kakazu et al. (2007)]{Kakazu07} Kakazu, Y., Cowie, L. L., 
\& Hu, E. M. 2007, \apj, 668, 853

\bibitem[Kewley \& Dopita (2002)]{KewleyDopita2002} Kewley, L. J.,
\& Dopita, M. A. 2002, \apjs,142, 35

\bibitem[Kewley et al. (2007)]{Kewley2007} Kewley, L. J., Brown, W. R., 
Geller, M. J., Kenyon, S. J., \& Kurtz, M. J. 2007, \aj, 133, 882

\bibitem[Kobulnicky \& Zaritsky (1999)]{KobylZarit1999} Kobulnicky, H. A., 
\& Zaritsky, D. 1999, \apj, 511, 118

\bibitem[Kobulnicky \& Koo (2000)]{KobKoo2000} Kobulnicky, H. A., \& Koo, D. C. 2000,
\apj, 545, 712

\bibitem[Kobulnicky \& Kewley (2004)]{KobKew04} Kobulnicky, H. A., 
\& Kewley, L. J. 2004, \apj, 617, 240  

\bibitem[Kobulnicky et al. (2003)]{Kobulniky2003} Kobulnicky, H. A.,
Willmer, C. N. A., Phillips, A. C., et al. 2003, \apj, 599, 1006

\bibitem[Kunth \& \"Ostlin (2000)]{KunthOstlin2000} Kunth, D., \& \"Ostlin, G.
2000, Astron Astrophys Rev, 10, 1 

\bibitem[Lamareille et al. (2004)]{Lamareille2004} Lamareille, F., Mouhcine, M., 
Contini, T., Lewis, L., \& Maddox, S. 2004, MNRAS, 350, 396

\bibitem[Lee et al. (2006)]{Lee45mu2006} Lee, H, Skillman, E. D., 
Cannon, J. M., et al. 2006, \apj, 647, 970

\bibitem[Lee et al. (2004)]{Lee2004}  Lee, J. C., Salzer, J. J., \&
Melbourne, J. 2004, \apj, 616, 752

\bibitem[Lequeux et al. (1979)]{Lequeux1979} Lequeux, J., Rayo, J. F., Serrano, A., 
Peimbert, M., \& Torres-Peimbert, S. 1979, \aap, 80, 155  

\bibitem[Liang et al. (2004)]{Liang2004} Liang, Y. C., Hammer, F., Flores, H., 
Elbaz, D., Marcillac, D., \& Cesarsky, C. J. 2004, \aap, 423, 867

\bibitem[Lilly et al. (2003)]{Lilly2003} Lilly, S. J., Carollo, C. M., 
\& Stockton, A. N. 2003, \apj, 597, 730

\bibitem[Maier et al. (2004)]{Maier2004} Maier, C., Meisenheimer, K., 
\& Hippelein, H. 2004, \aap, 418, 475

\bibitem[Melbourne \& Salzer (2002)]{MelbourneSalzer2002} Melbourne, J., \& 
Salzer, J. J.  2002, \aj, 123, 2302

\bibitem[Noeske et al. (2003)]{Noeske03}Noeske, K.G., Papaderos, P., Cairos,
  L.M., \& Fricke, K.J. 2003, \aap, 410, 481

\bibitem[Papaderos et al. (1996)]{P96} Papaderos, P., Loose, H.-H.,
Fricke, K. J., \& Thuan, T. X. 1996, \aap, 314, 59

\bibitem[Papaderos et al. (1998)]{P98} Papaderos, P., Izotov, Y. I.,
Fricke, K. J., Thuan, T. X., \& Guseva, N. G. 1998, \aap, 338, 43

\bibitem[Papaderos et al. (2002)]{P02} Papaderos, P., Izotov, Y. I., 
Thuan, T. X., Noeske, K. G., Fricke, K. J., Guseva, N. G., \& Green, R. F.
2002, \aap, 393, 461

\bibitem[Papaderos et al. (2008)]{Pap2008} Papaderos, P., Guseva, N. G., 
Izotov, Y. I., \& Fricke, K. J. 2008, \aap, 491, 113

\bibitem[Pettini et al. (2001)]{Pettini2001} Pettini, M., Shapley, A. E., 
Steidel, C. C., et al. 2001, \apj, 554, 981

\bibitem[Pilyugin et al. (2004)] {Pil2004} Pilyugin, L. S., V\'ilchez, J. M., 
\& Contini, T. 2004, \aap, 425, 849 

\bibitem[Pilyugin et al. (2006)]{Pil2006} Pilyugin, L. S., Thuan, T. X., \&
 V\'ilchez, J. M. 2007, \mnras, 367, 1139

\bibitem[Pilyugin et al. (2007)]{Pil2007} Pilyugin, L. S., Thuan, T. X., \&
 V\'ilchez, J. M. 2007, \mnras, 376, 353  

\bibitem[Richer \& McCall (1995)]{RicherMcC1995} Richer, M. G., 
\& MacCall, M. L. 1995, \apj, 445, 642

\bibitem[Salzer et al. (1989)]{Salzer1989} Salzer, J. J., MacAlpine, G. M.,
\& Boroson, T. A. 1989, \apjs, 70, 479
  
\bibitem[Salzer et al. (2005)]{Salzer05} Salzer, J. J., Lee, J. C., 
Melbourne, J., et al. 2005, \apj, 624, 661

\bibitem[Saviane et al. (2008)]{Saviane2008} Saviane, I., Ivanov, V. D., 
Held, E. V., et al. 2008, \aap, 487, 901

\bibitem[Shi et al. (2005)]{Shi2005} Shi, F., Kong, X., Li, C., \& Cheng, F. Z.
2005, \aap, 437, 849

\bibitem[Skillman et al. (1989)]{Skillman1989} Skillman, E. D., Kennicutt, R. C., Jr.
\& Hodge, P. W. 1989, \apj, 347, 875

\bibitem[Smith \& Hancock (2009)]{Smith2009} Smith, B. J., \& Hancock, M.
2009, \aj, 138, 130

\bibitem[Stasi\'nska (2002)]{Stas2002} Stasi\'nska, G. 2002, RMxAC, 12, 62

\bibitem[Stasi\'nska \& Izotov (2003)]{Stasin2003} Stasi\'nska, G., \& Izotov, Y. I. 
2003, \aap, 397, 71

\bibitem[Thuan \& Izotov (2005)]{TI2005} Thuan, T. X., \& Izotov, Y. I. 2005, 
\apjs, 161, 240

\bibitem[Tremonti et al. (2004)]{Tremonti2004} Tremonti, C. A., 
Heckman, T. M., Kauffmann, G., et al. 2004, \apj, 613, 898

\bibitem[Vanzi et al. (2000)]{Vanzi00}Vanzi, L., Hunt, L.K., Thuan, T.X., \&
  Izotov, Y.I. 2000, \aap, 363, 493

\bibitem[Vaduvescu et al. (2007)]{Vaduvescu07}Vaduvescu, O., McCall, M.L.,
  \& Richer, M.G. 2007, \aj, 134, 604

\bibitem[Vila-Costas \& Edmunds (1992)]{Vila1992} Vila-Costas, M. B.,  
\& Edmunds, M. G. 1992, MNRAS, 259, 121

\bibitem[Villar-Mart\'in et al. (2004)]{Villar2004} Villar-Mart\'in, M., 
Cervi\~no, M,  \& Gonz\'alez Delgado, R. M. 2004, MNRAS, 355, 1132

\bibitem[Whitford (1958)]{W58} Whitford, A. E. 1958, \aj, 63, 201

\bibitem[Zackrisson et al. (2008)]{Z08} Zackrisson, E., Bergvall, N.,\& Leitet,
  E. 2008, \apj, 676, L9

\end{thebibliography}
\end{document}